\documentclass[12pt]{article}
\usepackage{amsfonts}
\begin{document}
\newfont{\elevenmib}{cmmib10 scaled\magstep1}%
\newfont{\cmssbx}{cmssbx10 scaled\magstep3}
\newcommand{\preprint}{
            \begin{flushleft}
   \elevenmib Yukawa\, Institute\, Kyoto\\
            \end{flushleft}\vspace{-1.3cm}
            \begin{flushright}\normalsize  \sf
            YITP-98-76\\
   {\tt hep-th/9812232} \\ December 1998
            \end{flushright}}
\newcommand{\Title}[1]{{\baselineskip=26pt \begin{center}
            \Large   \bf #1 \\ \ \\ \end{center}}}
\newcommand{\Author}{\begin{center}\large \bf
            A.\, J.\, Bordner and
            R.\, Sasaki
            \end{center}}
\newcommand{\Address}{\begin{center} \it
             Yukawa Institute for Theoretical Physics, Kyoto
            University,\\ Kyoto 606-8502, Japan
      \end{center}}
\newcommand{\Accepted}[1]{\begin{center}{\large \sf #1}\\
            \vspace{1mm}{\small \sf Accepted for Publication}
            \end{center}}
\baselineskip=20pt

\preprint
\thispagestyle{empty}
\bigskip
\bigskip
\bigskip
\Title{Calogero-Moser Models III: Elliptic Potentials and
Twisting}
\Author

\Address
\vspace{2cm}

\begin{abstract}%
\noindent
Universal Lax pairs of the root type with
spectral parameter and  independent coupling constants for
twisted non-simply laced
 Calogero-Moser models are constructed. Together with the Lax
pairs for the simply laced models and untwisted non-simply
laced models presented in two previous papers, this completes
the derivation of universal Lax pairs for all of the
Calogero-Moser models based on root systems.
As for the twisted models based on \(B_{n}\), \(C_{n}\) and
\(BC_{n}\) root systems, a new type of potential term with
independent coupling constants can be added without destroying
integrability. They are called extended twisted models.
All of the Lax pairs for the twisted models presented here are new,
except for the one for the \(F_{4}\) model based on the short
roots. The Lax pairs for the twisted \(G_{2}\) model have some novel
features. Derivation of various functions, twisted and
untwisted, appearing in the Lax pairs for elliptic potentials
with the spectral parameter is provided.
The origin of the spectral parameter is also
naturally explained.
The Lax pairs with  spectral parameter, twisted and
untwisted, for the hyperbolic, the trigonometric and the
rational potential models are obtained as degenerate limits
of those for the elliptic potential models.
\\ \\
\end{abstract}

\newpage
\section{Introduction}
\setcounter{equation}{0}
This is the third paper in a series devoted to
the construction of universal Lax pairs for
Calogero-Moser models based on root systems. In
the first paper paper
\cite{bcs} (this paper will be referred to as I), a new and
universal formulation of Lax pairs of Calogero-Moser
models based on simply laced root systems was
presented. In the second paper \cite{bst}
(this paper will be referred to as II), the universal Lax
pairs for untwisted non-simply laced models with
independent coupling constants for the long and
short roots were constructed. The twisted non-simply
laced models were derived from the simply laced
models by folding with respect to a discrete symmetry of the
models.
This paper has two objectives:
firstly to complete the construction of universal Lax
pairs
of all Calogero-Moser models based on root systems
by constructing those for the twisted non-simply laced
models with independent coupling constants and
secondly to elucidate various functions appearing in
the Lax pairs with spectral parameter
\cite{bcs,bst}--\cite{DHPh} for the elliptic potential.
The Calogero-Moser models
\cite{CalMo} are a collection  of completely
integrable one-dimensional  dynamical systems
characterised by root systems and a choice of
four long-range  interaction potentials:
 (i) $1/L^2$, (ii) $1/\sin^2L$, (ii) $1/\sinh^2L$
and (iv) $\wp(L)$, (Weierstrass function) in which $L$
is the inter-particle ``distance''.
The hyperbolic, the trigonometric and the rational
potential models are obtained as degenerate limits of
the elliptic potential model.

For the general background and the motivations of this
series of papers and the physical applications of the
Calogero-Moser models with various potentials to
lower dimensional physics, ranging from solid state to
particle physics and  the Seiberg-Witten theory
\cite{SeiWit}, we refer to our previous papers I and II
and references therein.

We address  two problems in this paper:
the construction of universal Lax pairs with independent
coupling constants for twisted non-simply laced models
and a detailed exposition of the functions appearing in the
Lax pairs for the twisted and untwisted models.
Throughout this paper we focus on the elliptic potential
case with a spectral parameter.
The Weierstrass function \(\wp(u)\) is characterised by a
pair of primitive periods, which is denoted by
\(\{2\omega_1,2\omega_3\}\) (following the convention of
\cite{Erd}). As shown in paper II, the twisted models
have potentials \(\wp(u)\)  with the above periods
for the long roots and potentials \(\wp^{(1/n)}(u)\)
(\(n=3\) for the \(G_2\) model and \(n=2\) for the other
non-simply laced models) for the short roots, which have
\(1/n\)-th period in one direction
\(\{2\omega_1/n,2\omega_3\}\).

Our main result for the universal Lax pairs for twisted
non-simply laced models is very simple.
The Lax pairs for the twisted models can be obtained from those
for the untwisted models (given in paper II) by replacing the
untwisted functions
\(\{x,y,z,\ x_d,y_d,z_d,\ x_t,y_t,z_t\}\) by the twisted
functions \(\{x^{(1/n)},y^{(1/n)},z^{(1/n)},\
x^{(1/n)}_d,y^{(1/n)}_d,z^{(1/n)}_d,\
x^{(1/n)}_t,y^{(1/n)}_t,z^{(1/n)}_t\}\) whose explicit
forms will be given in section five.
All of the Lax pairs for the twisted models presented here are new,
except for the one for the \(F_{4}\) model based on the short roots.
The verification that the Lax equation is equivalent to the
canonical equation of motion and the consistency
of the Lax pairs is essentially the same as for the untwisted
models (paper II) and the simply laced models (paper I).
  As for
the twisted models based on
\(B_{n}\),
\(C_{n}\) and
\(BC_{n}\) root systems, a new type of potential terms with
independent coupling constants can be added without destroying
integrability.
They are called extended twisted models and they have three, three and
five independent coupling constants for \(B_{n}\), \(C_{n}\) and
\(BC_{n}\) models, respectively.
Some new features appear in the Lax pairs for the twisted \(G_{2}\)
model.

Our second objective is to give a detailed exposition of the
functions, twisted and untwisted, appearing in the Lax pairs.
The untwisted functions were used in previous papers I and II
without derivation.
Here we derive the explicit forms of these
functions starting from various sum rules
which are necessary and
sufficient for consistency of the Lax pairs for the elliptic
potential. The origin of the spectral parameter is also
explained. The Lax pairs with  spectral parameter, twisted and
untwisted, for the hyperbolic, the trigonometric and the
rational potential models are obtained as degenerate limits
of those for the elliptic potential models.

This paper is organised as follows.
In section two we recapitulate the essential ingredients of the
Calogero-Moser models with the elliptic potential.
 Section three is for the detailed exposition of the
untwisted functions.  Section four gives the root type Lax pairs
for all of the twisted models. Detailed accounts of the twisted
sum rules and functions are provided in section five. The final
section  is devoted to a summary and comments.

\section{Calogero-Moser models with elliptic potential}
\setcounter{equation}{0}
In order to set the stage and introduce notation
let us start with the definition of the Calogero-Moser
models based on a {\em simply laced} root system \(\Delta\),
which is
associated with a semi-simple and {\em simply-laced} Lie algebra
${\mathfrak g}$ of rank $r$.
The roots \(\alpha,\beta,\gamma,\ldots\/\) are real $r$-
dimensional vectors
and are normalised, without loss of generality, to 2:
\begin{equation}
    \Delta=\{\alpha,\beta,\gamma,\ldots\}, \quad \alpha\in
   {\bf R}^r,\quad
    \alpha^2=\alpha\cdot\alpha=2,\quad \forall\alpha\in\Delta.
    \label{eq:setroots}
\end{equation}

The dynamical variables are
 canonical coordinates $\{q^j\}$ and their
canonical conjugate momenta $\{p_j\}$ with the Poisson brackets:
\begin{equation}
    q^1,\ldots,q^r, \quad
   p_{1},\ldots,p_{r}, \quad
   \{q^j,p_{k}\}=\delta_{j,k},\quad \{q^j,q^{k}\}=\{p_{j},p_{k}\}=0.
    \label{eq:poisson}
\end{equation}
In most cases we denote them by $r$ dimensional vectors $q$ and $p$
\footnote{
For  $A_r$ models, it is customary to introduce one more degree of
freedom,
$q^{r+1}$ and $p_{r+1}$ and embed
all of the roots in ${\bf R}^{r+1}$.},
\begin{displaymath}
   q=(q^1,\ldots,q^r)\in {\bf R}^r,\quad
   p=(p_1,\ldots,p_r)\in {\bf R}^r,\quad
\end{displaymath}
so that the scalar products of $q$ and $p$ with the roots
$\alpha\cdot q$, $p\cdot\beta$, etc. can be defined.
The Hamiltonian with the elliptic potential is given by
\begin{equation}
   {\cal H}={1\over2}p^2+{g^2\over2}\sum_{\alpha\in\Delta}
   \wp(\alpha\cdot q),
   \label{eq:hamiltonianell}
\end{equation}
in which \(g\) is a (real) coupling constant. The Weierstrass
function
\(\wp\) is  a doubly periodic meromorphic function
with a pair of  primitive
periods
\(\{2\omega_1,2\omega_3\}\), \(\Im(\omega_3/\omega_1)>0\). We
adopt the convention
\cite{Erd} that the dependence on the  primitive
periods is suppressed for simplicity:
\begin{equation}
   \wp(u)\equiv\wp(u|
   2\omega_1,2\omega_3)={1\over{u^2}}+
   \sum_{m,\,n}{}^\prime\left[{1\over{(u-\Omega_{m,\,n})^2}}-
   {1\over{\Omega_{m,\,n}^2}}\right],
   \label{periods}
\end{equation}
in which \(\Omega_{m,\,n}\) is a period
\[
   \Omega_{m,\,n}=2m\omega_1+2n\omega_3
\]
and \(\sum{}^\prime\) denotes the summation over all
integers, positive, negative and zero, excluding
\(m=n=0\). Obviously the choice of the primitive
periods is not unique. Any {\em modular} (\(SL(2,{\bf Z})\))
transformation of  the pair of primitive periods is again
a pair of primitive periods:
\[
   \pmatrix{\omega_1^\prime\cr
         \omega_3^\prime\cr}
   =\pmatrix{a_{11}&a_{12}\cr
           a_{21}&a_{22}\cr}
   \pmatrix{\omega_1\cr
         \omega_3\cr},\quad a_{ij}\in{\bf Z}, \quad i,j=1,2,
   \quad a_{11}a_{22}-a_{12}a_{21}=1.
\]
Another
characterisation of Weierstrass function is  through
the invariants:
\[
   \wp(u)=\wp(u|2\omega_1,2\omega_3)=\wp(u|\{g_2,g_3\}),
\]
in which
\begin{equation}
   g_2=60\sum_{m,\,n}{}^\prime\,{\Omega_{m,\,n}^{-4}},
   \quad
   g_3=140\sum_{m,\,n}{}^\prime\,{\Omega_{m,\,n}^{-6}}.
   \label{ellinv}
\end{equation}
As is well known, the Weierstrass function \(\wp(z)\)
satisfies the differential equation
\begin{eqnarray}
   \left[\wp(u)^\prime\right]^2
   &=&4[\wp(u)]^3-g_2\wp(u)-g_3\nonumber\\
   &=& 4(\wp(u)-e_1)(\wp(u)-e_2)(\wp(u)-e_3),
   \label{weiinvchar}
\end{eqnarray}
in which
\begin{equation}
   e_1=\wp(\omega_1),\quad
   e_2=\wp(\omega_2),\quad
   e_3=\wp(\omega_3),\quad \omega_2\equiv -(\omega_1+\omega_3).
   \label{defeis}
\end{equation}

It should be noted that the above Hamiltonian is
real ({\em hermitian})  for real dynamical variables \(p,q\)
and coupling constant
\(g\), {\em provided}
\begin{equation}
   \omega_1: \mbox{real and}\quad
   \omega_3: \mbox{pure imaginary}.
   \label{hermcond}
\end{equation}
One can always choose a pair of primitive periods in this way
when
the two invariants \(g_2,g_3\) are
real and the {\em discriminant}
\begin{equation}
   Dis=g_3^2-27g_2^3
   \label{discrim}
\end{equation}
is {\em positive}.

\section{Functions in the Lax pairs for untwisted models}
\setcounter{equation}{0}
Let us now discuss the functions appearing in the
root type Lax pairs for the simply laced models and untwisted
non-simply laced models. These are the cases treated in
Paper I \cite{bcs} and II \cite{bst}. The {\em elliptic
potential case}  with spectral parameter will be discussed
in greater detail and the functions appearing in the other
cases, the rational, trigonometric and hyperbolic cases will
be obtained simply as the degeneration of those in the
elliptic case. The origin of the {\em spectral parameter} for
the elliptic case is naturally explained in
 our approach. It also leads, as a direct consequence of the
degeneration, to a not so widely recognised fact that
the spectral parameter can be introduced for the  rational,
trigonometric and hyperbolic cases, as well.

\bigskip

As shown in papers I and II, the Lax pair  for the
untwisted Calogero-Moser models contains functions
\(\{x(u),y(u),z(u)\}\), \(\{x_d(u),y_d(u),z_d(u)\}\)  and
\(\{x_t(u),y_t(u),z_t(u)\}\) (only in the untwisted
\(G_2\) model Lax pair based on long roots).
They  are required to satisfy the {\em first sum rule}:
\begin{equation}
   x^\prime(u)x(v)-x^\prime(v)x(u)=
   x(u+v)[\wp(v)-\wp(u)],
   \quad u,v\in {\bf C}.
   \label{eq:ident1}
\end{equation}
The same equations should hold for \(x_d\) and \(x_t\), too.
They all have a simple pole at \(u=0\) with unit residue:
\begin{equation}
   \lim_{u\to0}u\,x(u)=\lim_{u\to0}u\,x_d(u)=
   \lim_{u\to0}u\,x_t(u)=1.
   \label{rescond}
\end{equation}
The functions \(x\) and \(x_d\) are  required to satisfy
the {\em second sum rule} (see (I.2.29)):
\begin{eqnarray}
   x(u-v)\left[\wp(v)-\wp(u)\right]
   &+&2\left[x_d(u)\,y(-u-v)-
   y(u+v)\,x_d(-v)\right]\nonumber\\
   &+&x(u+v)\,y_d(-v)-y_d(u)\,x(-u-v)=0,
    \label{eq:xydiden}
\end{eqnarray}
and \(x\), \(x_d\) and \(x_t\) are  required to satisfy
the {\em third sum rule} (see (II.4.72)):
\begin{eqnarray}
   &&x(3u-3v)[\wp(2u-v)-\wp(u-2v)]
   -x(3v)\,y_t(u-2v)+y_t(2u-v)\,x(-3u)\nonumber\\
   &&-2x_d(3u)\,y_t(-u-v)+2y_t(u+v)\,x_d(-3v)
   -3x_t(2u-v)\,y(-3u)+3y(3v)\,x_t(u-2v)\nonumber\\
   &&-3x_t(u+v)\,y_d(-3v)+3y_d(3u)\,x_t(-u-v)=0,
   \label{tripiden}
\end{eqnarray}
in which
\begin{equation}
   y(u)\equiv x^\prime(u),\quad
   y_d(u)\equiv x_d^\prime(u),\quad y_t(u)\equiv x_t^\prime(u).
   \label{defys}
\end{equation}
The \(z\) functions are defined as the products of \(x\)
functions:
\begin{equation}
   z(u)\equiv x(u)x(-u),\quad
   z_d(u)\equiv
   x_d(u)x_d(-u),\quad
   z_t(u)\equiv
   x_t(u)x_t(-u).
   \label{defzs}
\end{equation}
It should be remarked that the set of solutions \(\{x(u),
x_d(u),x_t(u)\}\) to these sum rules has symmetry
transformations or a kind of `gauge freedom'. If
\(\{x(u), x_d(u),x_t(u)\}\) satisfies the first, the second and
the third sum rules, then
\begin{equation}
   \{ \tilde{x}(u)=x(u)e^{bu},\quad
   \tilde{x}_d(u)=x_d(u)e^{2bu},\quad
    \tilde{x}_t(u)=x_t(u)e^{3bu}\}
   \label{gaugefr}
\end{equation}
also satisfies the same sum rules. Here \(b\) is an arbitrary
\(u\)-independent constant, which can depend on the
spectral parameter \(\xi\). The functions \(z(u), z_d(u),
z_t(u)\) are invariant under these transformations.

It is interesting to note that the first (\ref{eq:ident1}),
second (\ref{eq:xydiden}) and third (\ref{tripiden}) sum
rules correspond to the three different types of rank 2 root
systems, \(A_2\), \(B_2 (C_2)\) and \(G_2\), respectively.
The rank 2 root systems are spanned by  the roots
\(\beta-\kappa\) and \(\kappa-\gamma\) in which \(\kappa\)
are non-trivial intermediate roots in the off-diagonal parts
of the commutator:
\begin{equation}
   [L,M]_{\beta,\,\gamma}
   =\sum_{\kappa\in\Delta}\left(L_{\beta,\,\kappa}
   M_{\kappa,\,\gamma}-M_{\beta,\,\kappa}
   L_{\kappa,\,\gamma}\right):
   \quad \beta, \gamma,\,\kappa\in\Delta.
\end{equation}
In fact, these three sum rules are obtained as the necessary
and sufficient conditions for the consistency of
the root type Lax pairs in the two-dimensional subspace
spanned by \(\beta-\kappa\) and \(\kappa-\gamma\).

\bigskip
Since the {\em first sum rule} (\ref{eq:ident1}) is the most
fundamental, let us discuss its general solution.
By taking the limit
\(v\to -u\) in (\ref{eq:ident1}) and using the residue
condition (\ref{rescond}) we obtain
\begin{equation}
   {d\over{du}}\{x(u)x(-u)\}=-{d\,\wp(u)\over{du}}.
   \label{eq:deriv}
\end{equation}
\noindent From this we know that
\begin{equation}
   x(u)x(-u)=-\wp(u)+const.
   \label{eq:integ}
\end{equation}
We can always choose this constant to be
\begin{equation}
   \wp(\xi),\quad \xi\in {\bf C}
   \label{eq:xidef}
\end{equation}
and obtain a factorised form \cite{Erd}:
\begin{equation}
  x(u)x(-u)=-\wp(u)+\wp(\xi)=-
  {\sigma(\xi-u)\over{\sigma(\xi)\sigma(u)}}
  {\sigma(\xi+u)\over{\sigma(\xi)\sigma(u)}}.
  \label{eq:facform}
\end{equation}
Here the Weierstrass sigma function \(\sigma(u)\) is
defined from
\(\wp(u)\) via the Weierstrass zeta function \(\zeta(u)\)
as:
\begin{eqnarray}
   \wp(u)&=&-\zeta^\prime(u),\qquad \quad
   \zeta(u)=d\log\,\sigma(u)/du=\sigma^\prime(u)/\sigma(u),
   \nonumber\\
   \zeta(u)&\equiv&\zeta(u|
   2\omega_1,2\omega_3)={1\over{u}}+
   \sum_{m,\,n}{}^\prime\left[{1\over{u-\Omega_{m,\,n}}}
   +{1\over{\Omega_{m,\,n}}}+{u\over{\Omega_{m,\,n}^2}}\right],
   \nonumber\\
   \sigma(u)&\equiv&\sigma(u|
   2\omega_1,2\omega_3)=u\,
   \prod_{m,\,n}{}^\prime\left(1-{u\over{\Omega_{m,\,n}}}\right)
   \exp\left[
   {u\over{\Omega_{m,\,n}}}+{u^2\over{2\Omega_{m,\,n}^2}}\right],
\end{eqnarray}
in which \(\prod{}^\prime\) denotes the product over all
integers, positive, negative and zero, excluding
\(m=n=0\). This is the simplest explanation of why the
{\em spectral parameter} \(\xi\) appears in the theory.
It is elementary to see that
\begin{equation}
   x(u)=x_{0}(u,\xi),\quad
   x_{0}(u,\xi)={\sigma(\xi-u)\over{\sigma(\xi)\sigma(u)}}
   \label{eq:elemsol}
\end{equation}
satisfies the {\em first sum rule} (\ref{eq:ident1}) for an
arbitrary constant
\(\xi\). By  the following monodromy property
of \( x_{0}(u,\xi)\):
\[
   x_0(u+2\omega_j,\xi)=e^{-2\eta_j\xi}\,x_{0}(u,\xi),\quad
   x_0(u,\xi+2\omega_j)=e^{-2\eta_j u}\,x_{0}(u,\xi), \quad
   \eta_j=\zeta(\omega_j),\quad j=1,2,3
\]
we find that
\[
   \left[x_0^\prime(u,\xi)x_0(v,\xi)-
   x_0^\prime(v,\xi)x_0(u,\xi)\right]/x_0(u+v,\xi)
\]
is a doubly periodic meromorphic function of
\(u,v\) and \(\xi\). Meromorphic functions are completely
determined by their zeros, poles and their residues.
It is trivial to check the zeros, poles and  residues
coincide with those of \(\wp(v)-\wp(u)\).
The other sum rules can be verified in a similar way.

\bigskip
Next we analyse possible poles and zeros
of \(x(u)\). From (\ref{eq:ident1}) it is clear that \(x(u)\)
cannot have a pole at \(u\neq0\) (modulo periods), because the
r.h.s. has no  singularity there. From the factorisation
(\ref{eq:facform}), it also follows that \(x(u)\)
cannot have a
zero at \(u\neq\{0,\pm\xi\}\).
This is because an additional zero in
\(x(u)\) must be canceled by a pole in
\(x(-u)\) which is not allowed
as above.
By the same argument, we find that
\(x(u)\) can have no other isolated singularities
(essential singularities or
branch points) at finite \(u\). Thus we deduce that \(x(u)\)
can be expressed as
\begin{equation}
   x(u)=x_{0}(u)e^{g(u)},\quad
   x_{0}(u)={\sigma(\xi-u)\over{\sigma(\xi)\sigma(u)}},\quad
   g(-u)=-g(u):\quad \mbox{entire function}.
   \label{eq:prodform}
\end{equation}
By substituting the above form in (\ref{eq:ident1}), we obtain
\begin{eqnarray}
    &  & x_{0}(u)x_{0}(v)\left(g'(u)-g'(v)\right)
   \nonumber  \\
    & = & x_{0}(u+v)\left(e^{g(u+v)-g(u)-g(v)}-1\right)
  \left(\wp(v)-\wp(u)\right).
   \label{eq:comp}
\end{eqnarray}
By comparing both sides at \(u=\infty\), we find the asymptotic
behaviour can only be matched for
\begin{equation}
   g(u)=\left\{
   \begin{array}{l}
    0\\
    b\,u \quad b: \quad const.
   \end{array}
   \right.
   \label{eq:sol}
\end{equation}
It should be mentioned that \(x_{0}(u)\) is of
exponential growth,
i.e., \(x_{0}(u)\sim e^{cu}\) for some constant \(c\).
The constant \(b\) can depend on \(\xi\). Thus we arrive at
the conclusion that the general solution of the
{\em first sum rule} (\ref{eq:ident1}) is given by
\begin{equation}
   x(u)={\sigma(\xi-u)\over{\sigma(\xi)\sigma(u)}}\,e^{bu},
   \quad \xi,\, b: const.
   \label{gensol}
\end{equation}
A similar but less direct analysis of solutions of a
functional equation which is a certain
generalisation of the first sum rule has been given in
\cite{BrCal}.

Since the functions \(x_d\) and \(x_t\) must be of the above
general form (\ref{gensol}), it is easy to see that the
general solutions of the {\em first, second} and  {\em
third sum rules}
are given by (see (II.2.20), (II.2.21), (II.4.74)):
\begin{equation}
   x(u)={\sigma(\xi-u)\over{\sigma(\xi)\sigma(u)}}\,e^{bu},
   \quad
   x_d(u)={\sigma(2\xi-u)\over{\sigma(2\xi)\sigma(u)}}\,e^{2bu},
   \quad
   x_t(u)={\sigma(3\xi-u)\over{\sigma(3\xi)\sigma(u)}}\,e^{3bu}.
   \label{threesol}
\end{equation}
This solution also gives the same \(z, z_d,z_t\) (\ref{defzs}) up
to constants:
\begin{equation}
   z(u)-z_d(u)=\wp(\xi)-\wp(2\xi),\quad
   z(u)-z_t(u)=\wp(\xi)-\wp(3\xi).
   \label{samezs}
\end{equation}
It is easy to derive the following relation
\begin{equation}
   x(2u)x_d(-u)+x(-2u)x_d(u)=-\wp(u)+\wp(\xi)
   \label{secfac}
\end{equation}
\noindent from the second sum rule (\ref{eq:xydiden}). One only
has to follow the same path as the factorisation identity
obtained from the first sum rule. That is, to take the limit
\(v\to u\) in (\ref{eq:xydiden})
\[
   {d\over{du}}\left(x(2u)x_d(-u)+x(-2u)x_d(u)\right)=
   -{d\over{du}}\wp(u)
\]
and integrate it. The above equation and (\ref{secfac})
are responsible for the ``renormalisation" of the short root
coupling constant in the untwisted \(BC_n\) model (II.4.85).
In a similar way one obtains the following relation from the
third sum rule (\ref{tripiden}):
\begin{equation}
   x(3u)x_t(-u)+x(-3u)x_t(u)+x_d(3u)x_t(-2u)+x_d(-3u)x_t(2u)
   =-\wp(u)+\wp(\xi).
   \label{thirdfac}
\end{equation}

\bigskip
At the end of this subsection, let us remark that one can
rewrite the first sum rule in an equivalent form
\begin{equation}
   x^\prime(u)x(v)-x^\prime(v)x(u)=
   x(u+v)[x(u)x(-u) -x(v)x(-v)],
   \label{eq:ident1equiv}
\end{equation}
which can be read as an {\em addition theorem}:
\begin{equation}
   x(u+v)=(x^\prime(u)x(v)-x^\prime(v)x(u))/[x(u)x(-u)
   -x(v)x(-v)].
   \label{addition}
\end{equation}

\subsection{Degenerate cases}
\label{untwistdeg}
It is well known that the degenerates cases of Weierstrass
functions occur when one or both of the periods become
infinite.
The degenerate cases occur if two or all three of \(e_1,
e_2, e_3\) (\ref{defeis}) coincide \cite{Erd}.
There are three cases, corresponding to the hyperbolic,
trigonometric and
rational potential case, respectively. The degenerate
limit of the functions \(\{x, x_d, x_t\}\) in
(\ref{threesol}) gives the corresponding Lax pair
with {\em spectral parameter},  which is a new result.

\subsubsection{Real period infinite: Hyperbolic potential}
In this case \cite{Erd}
\begin{equation}
   e_1=e_2=a\in{\bf R_+},\quad e_3=-2a,\quad
   g_2=12a^2,\quad g_3=-8a^3,\quad
   \omega_1=\infty,\quad \omega_3={\pi i\over\sqrt{12a}}
   \label{realinf}
\end{equation}
and
\begin{eqnarray}
   \wp(u|\{12a^2,-8a^3\})&=&a+\kappa^2{1\over{\sinh^2 \kappa
   u}},\quad \kappa=\sqrt{3a},\nonumber\\
   \zeta(u|\{12a^2,-8a^3\})&=&-au+\kappa\coth \kappa
   u,\nonumber\\
   \sigma(u|\{12a^2,-8a^3\})&=&\sinh[\kappa
   u]\,\exp[-au^2/2]/\kappa.\label{realinffun}
\end{eqnarray}
The Calogero-Moser model has a hyperbolic
potential. The functions appearing in the Lax pair
for the elliptic potential with spectral parameter \(\xi\)
\begin{eqnarray}
   x(u,\xi)&\equiv&{\sigma(\xi-u)\over{\sigma(\xi)\sigma(u)}},
   \qquad \quad
   x_d(u,\xi)\equiv{\sigma(2\xi-u)\over{\sigma(2\xi)\sigma(u)}}
   =x(u,2\xi),
   \nonumber\\
   x_t(u,\xi)&\equiv&{\sigma(3\xi-u)\over{\sigma(3\xi)\sigma(u)}}
   =x(u,3\xi),
\label{untwiste-sols}
\end{eqnarray}
reduce to
\begin{eqnarray}
   x(u,\xi)&=&\kappa\left(\coth \kappa u-\coth \kappa
   \xi\right)\,e^{a\xi u},
   \qquad
   x_d(u,\xi)=\kappa\left(\coth \kappa u-\coth 2\kappa
   \xi\right)\,e^{2a\xi u},
   \nonumber\\
   x_t(u,\xi)&=&\kappa\left(\coth \kappa u-\coth 3\kappa
   \xi\right)\,e^{3a\xi u}.
   \label{realinffunx}
\end{eqnarray}
By utilising the ``gauge freedom" (\ref{gaugefr}), the
exponential part of the above functions can be removed
to obtain  simple spectral parameter dependent functions:
\begin{eqnarray}
   x(u,\xi)&=&\kappa(\coth \kappa u-\coth \kappa
   \xi),
   \qquad
   x_d(u,\xi)=\kappa(\coth \kappa u-\coth 2\kappa
   \xi),\nonumber\\
   x_t(u,\xi)&=&\kappa(\coth \kappa u-\coth 3\kappa
   \xi).
   \label{cothsol}
\end{eqnarray}
If we take the limit \({\kappa\xi\to\pm\infty}\),
we obtain
the spectral parameter independent functions
\begin{equation}
   x(u)=x_d(u)=x_t(u)=a\,(\coth a u\mp1),\quad a\,:\ const,
   \label{cothind}
\end{equation}
which are essentially the same as the spectral parameter
independent functions given in previous papers
(I.2.7), (II.2.13), (II.4.73).

\subsubsection{Imaginary period infinite: Trigonometric
potential}
In this case
\begin{equation}
   e_1=2a\in{\bf R_+},\quad e_2=e_3=-a,\quad
   g_2=12a^2,\quad g_3=8a^3,\quad
   \omega_1={\pi\over\sqrt{12a}},\quad \omega_3=i\infty
   \label{imainf}
\end{equation}
and
\begin{eqnarray}
   \wp(u|\{12a^2,8a^3\})&=&-a+\kappa^2{1\over{\sin^2 \kappa
   u}},\quad \kappa=\sqrt{3a},\nonumber\\
   \zeta(u|\{12a^2,8a^3\})&=&au+\kappa\cot \kappa
   u,\nonumber\\
   \sigma(u|\{12a^2,8a^3\})&=&\sin[\kappa
   u]\,\exp[au^2/2]/\kappa.\label{imainffun}
\end{eqnarray}
The corresponding Calogero-Moser model has a trigonometric
potential. The functions appearing in the Lax pair
for the elliptic potential with spectral parameter \(\xi\)
(\ref{untwiste-sols})
reduce to
\begin{eqnarray}
   x(u,\xi)&=&\kappa\left(\cot \kappa u-\cot \kappa
   \xi\right)\,e^{-a\xi u},
   \qquad
   x_d(u,\xi)=\kappa\left(\cot \kappa u-\cot 2\kappa
   \xi\right)\,e^{-2a\xi u},
   \nonumber\\
   x_t(u,\xi)&=&\kappa\left(\cot \kappa u-\cot 3\kappa
   \xi\right)\,e^{-3a\xi u}.
   \label{imainffunx}
\end{eqnarray}
As above the the
exponential part of the above functions can be removed
by using the ``gauge freedom" of (\ref{gaugefr}):
\begin{eqnarray}
   x(u,\xi)&=&\kappa(\cot \kappa u-\cot \kappa
   \xi),
   \qquad
   x_d(u,\xi)=\kappa(\cot \kappa u-\cot 2\kappa
   \xi),\nonumber\\
   x_t(u,\xi)&=&\kappa(\cot \kappa u-\cot 3\kappa
   \xi).
   \label{cotsol}
\end{eqnarray}
If we take the limit \({\kappa\xi\to\pm i\infty}\), we obtain
the spectral parameter independent functions
\begin{equation}
   x(u)=x_d(u)=x_t(u)=a\,(\cot a u\pm i),\quad a\,:\ const,
\end{equation}
which are essentially the same as the spectral parameter
independent functions given in previous papers
(I.2.6), (II.2.12), (II.4.73).

\subsubsection{Both periods infinite: Rational
potential}
This case is very simple:
\begin{equation}
   e_1=e_2=e_3=0,\quad
   g_2=g_3=0,\quad
   \omega_1=-i\omega_3=\infty
   \label{bothinf}
\end{equation}
and
\begin{equation}
   \wp(u|\{0,0\})=u^{-2},\quad
   \zeta(u|\{0,0\})=u^{-1},\quad
   \sigma(u|\{0,0\})=u.
   \label{bothinffun}
\end{equation}
In this case the Calogero-Moser model has the rational
potential and the functions appearing in the Lax pair
for the elliptic potential with spectral parameter \(\xi\)
(\ref{untwiste-sols})
reduce to
\begin{equation}
   x(u,\xi)={1\over{u}}-{1\over{\xi}},\quad
   x_d(u,\xi)={1\over{u}}-{1\over{2\xi}},\quad
   x_t(u,\xi)={1\over{u}}-{1\over{3\xi}}.
   \label{ratspec}
\end{equation}
Thus the Lax pair for the rational potential can also have
a spectral parameter, although the dependence on it
 is very simple. By taking the limit as the spectral
parameter goes to infinity, we recover the well known spectral
parameter independent functions
\begin{equation}
   x(u)=x_d(u)=x_t(u)={1\over{u}},
   \label{ratindep}
\end{equation}
which have been used in our previous papers
(I.2.5), (II.2.11), (II.4.73).

In the above three degenerate cases, the discriminant
(\ref{discrim}) is zero.

\section{Universal Lax pairs for Calogero-Moser models based
on twisted
non-simply laced algebras}
\setcounter{equation}{0}

In this section we present universal Lax pairs for
twisted Calogero-Moser models with independent coupling
constants. In section 3 of paper II, the twisted
Calogero-Moser models are derived from the simply laced
models with the elliptic potential by folding with respect
to a discrete symmetry
originating from the automorphism
of the extended Dynkin diagrams  combined with the
periodicity of the potential.
The integrability of these reduced models is inherited from
the original simply laced models.  In these
reduced models, however, the coupling  constants for the
long and short root potentials have a fixed ratio,
since the simply laced models  have only one coupling.
In order to exhibit the full symmetry
of the twisted non-simply laced models, we construct
root type Lax pairs with independent coupling constants.
As in the untwisted non-simply laced models,
there are two kinds of root type Lax pairs for the
twisted non-simply laced models, the one based on long roots
and the other short roots. Both are relatively  straightforward
generalisations of the root type Lax pairs for untwisted
non-simply laced systems.
Our main result for the universal Lax pairs for twisted
non-simply laced models is very simple.
The Lax pairs for the twisted models can be obtained from
those for the untwisted models (given in paper II)
by replacing the untwisted functions
\(\{x,y,z,\ x_d,y_d,z_d,\ x_t,y_t,z_t\}\) by the twisted
functions \(\{x^{(1/n)},y^{(1/n)},z^{(1/n)},\
x^{(1/n)}_d,y^{(1/n)}_d,z^{(1/n)}_d,\
x^{(1/n)}_t,y^{(1/n)}_t,z^{(1/n)}_t\}\) whose explicit
forms will be given in section five.
Among the twisted models, \(C_n\),
\(B_n\) and \(BC_n\) models are different from the rest, the \(F_4\)
and \(G_2\) models.
In the former models, the long roots of  \(C_n\) (\(\pm 2e_j,
j=1,\ldots,n\)), the short roots of \(B_n\) (\(\pm e_j,
j=1,\ldots,n\)) and the long and short roots of \(BC_n\)
(\(\pm 2e_j,\pm e_j,
j=1,\ldots,n\)) share one special property, that is, they are
orthogonal to each other:
\begin{equation}
\alpha\cdot\beta=0,\quad \alpha\neq\{\pm \beta,\quad \pm
{1\over2}\beta,
\quad \pm 2\beta\}.
\label{orthog}
\end{equation}
This fact enables us to introduce another independent coupling constant
in
twisted \(C_n\) and \(B_n\) models and two more independent couplings
in the \(BC_n\) model.
Let us call these models with extra independent coupling constant(s)
the {\em extended twisted models.}

Except for the Lax pair for the twisted \(F_4\) model based on short roots
\cite{DHPh} treated in \ref{f4short},
all the Lax pairs reported in this section are new.
The verification that the Lax equation is equivalent to the
canonical equation of motion and the consistency
of the Lax pairs is essentially the same as in the untwisted
models (paper II) and the simply laced models (paper I).

\subsection{Extended twisted \(C_n\) model}
The set of \(C_n\) roots consists of two parts,
long roots and short roots:
\begin{equation}
   \Delta_{C_n}=\Delta_l\cup\Delta_s,
\end{equation}
in which the roots are conveniently expressed in terms of
an orthonormal basis of \({\bf R}^n\):
\begin{eqnarray}
   \Delta_l&=&\{\Xi,\Upsilon,\Omega,\ldots,\}=\{\pm 2e_j:
   \qquad \quad j=1,\ldots,n\},
   \quad 2n\ \mbox{roots},
   \label{cnLnroots}\\
   \Delta_s&=&\{\alpha,\beta,\gamma,\ldots, \}=
   \{\pm e_j\pm e_k: \quad j\neq k=1,\ldots,n\},\quad
   2n(n-1)\ \mbox{roots}.
   \label{cnshroots}
\end{eqnarray}
 The Hamiltonian of {\em untwisted} \(C_n\) model is given by
\begin{equation}
    {\cal H}_{C_n}^{untwisted}={1\over2}p^2+{g_s^2\over2}
   \sum_{\alpha\in\Delta_s}
   \wp(\alpha\cdot q)+{g_l^2\over4}\sum_{\Xi\in\Delta_l}
    \wp(\Xi\cdot q),
    \label{eq:untwistcngenham}
\end{equation}
in which \(g_l\) and \(g_s\) are the two coupling constants
for the long roots and short roots. The twisted model is
obtained by multiplying one of the primitive periods of the
potential for the short roots by a factor of one half (II.3.21):
\begin{equation}
    {\cal H}_{C_n}^{twisted}={1\over2}p^2+{g_s^2\over2}
   \sum_{\alpha\in\Delta_s}
   \wp^{(1/2)}(\alpha\cdot q)+{g_l^2\over4}\sum_{\Xi\in\Delta_l}
    \wp(\Xi\cdot q).
    \label{eq:twistcngenham}
\end{equation}
In this paper we choose \(\{2\omega_1,2\omega_3\}\ \to
\{\omega_1,2\omega_3\}\), that is the potential for the short
roots is
\begin{equation}
   \wp^{(1/2)}(u)\equiv\wp(u|
   \omega_1,2\omega_3)=\wp(u)+\wp(u+\omega_1)-\wp(\omega_1).
   \label{halfperiods}
\end{equation}
The second equality is known as Landen's transformation.
The other case \(\{2\omega_1,2\omega_3\}\ \to
\{2\omega_1,\omega_3\}\) can be treated in a similar way.
The extended twisted \(C_n\) model is  defined, roughly speaking,
by a sum of the untwisted Hamiltonian at a half-period and the
twisted Hamiltonian:
\begin{eqnarray}
 {\cal H}_{C_n}^{ex-twisted}&=&{{\cal H}_{C_n}^{untwisted}}^{(1/2)}+
   {\cal H}_{C_n}^{twisted}\nonumber\\
   &=&{1\over2}p^2+{g_s^2\over2}
   \sum_{\alpha\in\Delta_s}
   \wp^{(1/2)}(\alpha\cdot q)+{1\over4}\sum_{\Xi\in\Delta_l}
   \left[\tilde{g}_{l_1}^2\wp(\Xi\cdot q)
   +g_{l_2}^2\wp^{(1/2)}(\Xi\cdot q)
   \right],
  \label{excnHam}
\end{eqnarray}
in which the ``renormalised" long root coupling constant is defined by
\begin{equation}
\tilde{g}_{l_1}^2=g_{l_1}(g_{l_1}+2g_{l_2})
\label{long-renom}
\end{equation}
from the parameters in the Lax pairs (\ref{eq:cnlnXYLdef}) and
(\ref{eq:cnlnXYdef}).
The potential for the long root has two terms, the Weierstrass function
with periods \(\{2\omega_1,2\omega_3\}\) and the one with
the half-period \(\{\omega_1,2\omega_3\}\). Obviously the
above Hamiltonian (\ref{excnHam}) is integrable for
\(\tilde{g}_{l_1}^2=0\) or
\(g_{l_2}^2=0\). The orthogonality of the long roots (\ref{orthog})
allows these two  potentials to co-exist
without breaking the integrability.

\bigskip
The  Lax pairs for the twisted model are simply obtained from
those of the untwisted model by changing  the functions \(x,
x_d\) and \(y, y_d\) corresponding to short roots to new ones
having a half-period.

\subsubsection{Root type Lax pair for extended twisted \(C_n\) model
based on short roots \(\Delta_s\)}
\label{cnshort}
The Lax pair is given in terms of short roots. It is
determined by  the root difference pattern of
the short roots:
\begin{equation}
   C_n:\qquad \mbox{short root}
   - \mbox{short root}=\left\{
   \begin{array}{l}
      \mbox{short root}\\
      2\times \mbox{short root}\\
      \mbox{long root}\\
      \mbox{non-root}
   \end{array}
   \right.
   \label{cnshrtshrt}
\end{equation}
The matrix elements of \(L_s\) and \(M_s\)  are
labeled  by indices
\(\beta,\gamma\) etc.:
\begin{eqnarray}
    L_s(q,p,\xi) & = & p\cdot H + X + X_{d}+X_L,
   \nonumber\\
    M_s(q,\xi) & = & D+D_L+Y+Y_{d}+X_L,
    \label{eq:cnshLaxform}
\end{eqnarray}
Here \(X\) and \(Y\) correspond to the part
of ``short root $-$ short root $=$ short root'' of
(\ref{cnshrtshrt}):
\begin{equation}
    X=ig_s\sum_{\alpha\in\Delta_s}x^{(1/2)}(\alpha\cdot
    q, \xi)E(\alpha),\quad
    Y=ig_s\sum_{\alpha\in\Delta_s}y^{(1/2)}(\alpha\cdot
    q, \xi)E(\alpha),\quad
    E(\alpha)_{\beta \gamma}=\delta_{\beta-\gamma,\alpha},
    \label{eq:cnshXYdef}
\end{equation}
and $X_d$ and $Y_d$ correspond to
``short root $-$ short root $=$ 2\(\times\) short
root'' of (\ref{cnshrtshrt}):
\begin{equation}
    X_d=2ig_s\sum_{\alpha\in\Delta_s}
    x_{d}^{(1/2)}(\alpha\cdot q, \xi)E_{d}(\alpha),\quad
    Y_d=ig_s\sum_{\alpha\in\Delta_s}
    y_{d}^{(1/2)}(\alpha\cdot q, \xi)E_{d}(\alpha),\quad
    E_{d}(\alpha)_{\beta \gamma}=\delta_{\beta-\gamma,2\alpha}.
    \label{eq:cnshXYrdef}
\end{equation}
The terms in \(L_s\) (\(M_s\)),
$X_L$ ($Y_L$) corresponding to
``short root $-$ short root $=$ long root''
of (\ref{cnlonglong}) have two terms corresponding to two independent
coupling constants  for the long root potentials:
\begin{eqnarray}
    X_L&=&i\sum_{\Xi\in\Delta_l}
    \left[g_{l_1}x(\Xi\cdot q, \xi)+g_{l_2}x^{(1/2)}(\Xi\cdot q,
    \xi)\right]E(\Xi),\nonumber\\
    Y_L&=&i\sum_{\Xi\in\Delta_l}
    \left[g_{l_1}y(\Xi\cdot q, \xi)+g_{l_2}y^{(1/2)}(\Xi\cdot q,
    \xi)\right]E(\Xi),\nonumber\\
    E(\Xi)_{\beta
\gamma}&=&\delta_{\beta-\gamma,\Xi}.
    \label{eq:cnlnXYLdef}
\end{eqnarray}
The diagonal parts of \(L_s\) and \(M_s\) are given by
\begin{equation}
    H_{\beta \gamma}=\beta \delta_{\beta, \gamma},\quad
    D_{\beta \gamma}= \delta_{\beta, \gamma}D_{\beta},\quad
    D_{\beta}=-ig_s\left(z^{(1/2)}(\beta\cdot q)
   +\sum_{\kappa\in\Delta_s,\
    \kappa\cdot\beta=1}z^{(1/2)}(\kappa\cdot q)\right),
   \label{eq:cnshHD}
\end{equation}
and
\begin{equation}
   (D_L)_{\beta \gamma}= \delta_{\beta,
   \gamma}(D_L)_{\beta},\quad
    (D_L)_{\beta}=-i\sum_{\Upsilon\in\Delta_l,\
    \beta\cdot\Upsilon=2}\left[g_{l_1}z(\Upsilon\cdot q)
    +g_{l_2}z^{(1/2)}(\Upsilon\cdot q)\right].
    \label{eq:cnshdLHD}
\end{equation}
The new functions
\(x^{(1/2)},y^{(1/2)}={x^{(1/2)}}^\prime,z^{(1/2)}\) and
\(x_d^{(1/2)},y_d^{(1/2)}={x_d^{(1/2)}}^\prime\) must satisfy
certain sum rules which are generalisations of the second sum
rule. These sum rules are listed in section~\ref{twfuns}
together with the explicit forms of the new functions.
It is easy to verify
\[
   Tr(L_s^2)=8(n-1){\cal H}_{C_n}^{ex-twisted},
\]
in which a factorisation identity (\ref{fac11n}) is
responsible for the renormalisation of the long root
coupling.

\subsubsection{Root type Lax pair for extended twisted
\(C_n\) model based on long roots
\(\Delta_l\)}
\label{cnlong}
The Lax pair is given in terms of long roots.
It is
determined by  the root difference pattern of
the long roots:
\begin{equation}
   C_n:\qquad \mbox{long root}
   - \mbox{long root}=\left\{
   \begin{array}{l}
      2\times \mbox{long root}\\
      2\times \mbox{short root}\\
      \mbox{non-root}
   \end{array}
   \right.
   \label{cnlonglong}
\end{equation}
The matrix elements of \(L_L\) and \(M_L\)  are
labeled  by indices
 \(\Upsilon,\Omega\)
etc.:
\begin{eqnarray}
    L_L(q,p,\xi) & = & p\cdot H +  X_{d}+X_s, \nonumber\\
    M_L(q,\xi) & = & D+Y_{d}+Y_s.
    \label{eq:cnlnLaxform}
\end{eqnarray}
Here \(X_d\) and \(Y_d\) correspond to the part
of ``long root $-$ long root $=2\times$ long root'' of
(\ref{cnlonglong}):
\begin{eqnarray}
    X_d&=&2i\sum_{\Xi\in\Delta_l}\left[g_{l_1}x_d(\Xi\cdot
   q, \xi)+g_{l_2}x_d^{(1/2)}(\Xi\cdot
   q, \xi)\right]E_d(\Xi),\nonumber\\
    Y_d&=&i\sum_{\Xi\in\Delta_l}\left[g_{l_1}y_d(\Xi\cdot
   q, \xi)+g_{l_2}y_d^{(1/2)}(\Xi\cdot
   q, \xi)\right]E_d(\Xi),\nonumber\\
    E_d(\Xi)_{\Upsilon \Omega}&=&\delta_{\Upsilon-\Omega,2\Xi},
    \label{eq:cnlnXYdef}
\end{eqnarray}
and $X_s$ and $Y_s$ correspond to
``long root $-$ long root $=$ 2\(\times\) short root''
of (\ref{cnlonglong}):
\begin{equation}
    X_s=2ig_s\sum_{\alpha\in\Delta_s}
   x_{d}^{(1/2)}(\alpha\cdot q, \xi)E_{d}(\alpha),\quad
    Y_s=ig_s\sum_{\alpha\in\Delta_s}
   y_{d}^{(1/2)}(\alpha\cdot q, \xi)E_{d}(\alpha),\quad
    E_{d}(\alpha)_{\Upsilon
   \Omega}=\delta_{\Upsilon-\Omega,2\alpha}.
    \label{eq:cnlnXYrdef}
\end{equation}
The diagonal parts of \(L_L\) and \(M_L\) are given by
\begin{eqnarray}
    H_{\Upsilon \Omega}&=&\Upsilon \delta_{\Upsilon, \Omega},\quad
    D_{\Upsilon \Omega}= \delta_{\Upsilon,
   \Omega}D_{\Upsilon},\nonumber\\
    D_{\Upsilon}&=&-i\left(g_{l_1}z_d(\Upsilon\cdot
   q, \xi)+g_{l_2}z_d^{(1/2)}(\Upsilon\cdot
   q, \xi)+g_s\sum_{\kappa\in\Delta_s,\
    \kappa\cdot\Upsilon=2}z_d^{(1/2)}(\kappa\cdot q)\right).
   \label{eq:cnlnHD}
\end{eqnarray}
The  functions
\(x^{(1/2)},y^{(1/2)}={x^{(1/2)}}^\prime,z^{(1/2)}\) and
\(x_d^{(1/2)},y_d^{(1/2)}={x_d^{(1/2)}}^\prime\) are the
same as those appearing in the Lax pair based on short roots.
Needless to say, the
consistency of the root type Lax pairs
(\ref{eq:cnshLaxform}) and  (\ref{eq:cnlnLaxform})
does not depend on the explicit representation of the
roots in terms of the orthonormal basis
(\ref{cnLnroots},\ref{cnshroots}). This remark applies to the
other models as well.

\subsection{Extended twisted \(B_n\) model}
The set of \(B_n\) roots consists of two parts,
long roots and short roots:
\begin{equation}
   \Delta_{B_n}=\Delta_l\cup\Delta_s,
\end{equation}
in which the roots are conveniently expressed in terms of
an orthonormal basis of \({\bf R}^n\):
\begin{eqnarray}
   \Delta_l&=&\{\alpha,\beta,\gamma,\ldots,\}=\{\pm e_j\pm
   e_k:
   \quad j\neq k=1,\ldots,n\},
   \quad 2n(n-1)\ \mbox{roots},
   \nonumber\\
   \Delta_s&=&\{\lambda,\mu,\nu,\ldots, \}=\{\pm e_j: \qquad \quad
   j=1,\ldots,n\},
   \quad 2n\ \mbox{roots.}
   \label{bnroots}
\end{eqnarray}
 The Hamiltonian of the {\em untwisted}  \(B_n\) model
is given by
\begin{equation}
    {\cal H}_{B_n}^{untwisted}={1\over2}p^2+{g_l^2\over2}
   \sum_{\alpha\in\Delta_l}
    \wp(\alpha\cdot q)+
   {g_s^2}\sum_{\lambda\in\Delta_s}
    \wp(\lambda\cdot q).
    \label{eq:untwistbngenham}
\end{equation}
As in the \(C_n\) case, the {\em twisted} model is
obtained by multiplying one of the primitive periods of the
potential for the short roots by a factor of one half (II.3.33):
\begin{equation}
    {\cal H}_{B_n}^{twisted}={1\over2}p^2+{g_l^2\over2}
   \sum_{\alpha\in\Delta_l}
    \wp(\alpha\cdot q)+
   {g_s^2}\sum_{\lambda\in\Delta_s}
    \wp^{(1/2)}(\lambda\cdot q).
    \label{eq:twistbngenham}
\end{equation}
The extended twisted \(B_n\) model is  defined, roughly speaking,
as a sum of the untwisted Hamiltonian  and the
twisted Hamiltonian:
\begin{eqnarray}
 {\cal H}_{B_n}^{ex-twisted}&=&{{\cal H}_{B_n}^{untwisted}}+
   {\cal H}_{B_n}^{twisted}\nonumber\\
   &=&{1\over2}p^2+{g_l^2\over2}
   \sum_{\alpha\in\Delta_l}
   \wp(\alpha\cdot q)+\sum_{\lambda\in\Delta_s}
   \left[\tilde{g}_{s_1}^2\wp(\lambda\cdot
   q)+g_{s_2}^2\wp^{(1/2)}(\lambda\cdot q)
   \right],
  \label{exbnHam}
\end{eqnarray}
in which the renormalised short root coupling constant is
defined by
\begin{equation}
   \tilde{g}_{s_1}^2=g_{s_1}(g_{s_1}+2g_{s_2})
   \label{short-renom}
\end{equation}
from the parameters in the Lax pairs (\ref{eq:bnshXYrdef}) and
(\ref{eq:bnlnXYsdef}).

\subsubsection{Root type Lax pair for extended twisted \(B_n\)
model based on short roots
\(\Delta_s\)}
\label{bnlong}
The Lax pair is given in terms of the short roots. %
It is
determined by  the root difference pattern of
the short roots:
\begin{equation}
   B_n:\qquad \mbox{short root}
   - \mbox{short root}=\left\{
   \begin{array}{l}
      \mbox{long root}\\
      2\times \mbox{short root}\\
      \mbox{non-root}
   \end{array}
   \right.
   \label{bnshrtshrt}
\end{equation}
The matrix elements of \(L_s\) and \(M_s\)  are
labeled  by indices
\(\mu,\nu\) etc.:
\begin{eqnarray}
    L_s(q,p,\xi) & = & p\cdot H + X + X_{d},\quad
    \nonumber\\
    M_s(q,\xi) & = & D+Y+Y_{d}.
    \label{eq:bnshLaxform}
\end{eqnarray}
Here \(X\) and \(Y\) correspond to the part
of ``short root $-$ short root $=$ long root'' of
(\ref{bnshrtshrt}):
\begin{equation}
    X=ig_l\sum_{\alpha\in\Delta_l}x(\alpha\cdot
    q, \xi)E(\alpha),\quad
    Y=ig_l\sum_{\alpha\in\Delta_l}y(\alpha\cdot
    q, \xi)E(\alpha),\quad
    E(\alpha)_{\mu \nu}=\delta_{\mu-\nu,\alpha}.
    \label{eq:bnshXYdef}
\end{equation}
Thus they have the same form as in the untwisted model.
The difference occurs in $X_d$ and $Y_d$, which correspond to
``short root $-$ short root $=$ 2\(\times\) short
root'' of (\ref{bnshrtshrt}):
\begin{eqnarray}
    X_d&&=2i\sum_{\lambda\in\Delta_s}\left[
    g_{s_1}x_{d}(\lambda\cdot q,\xi)+
    g_{s_2}x_{d}^{(1/2)}(\lambda\cdot q, \xi)\right]E_{d}(\lambda),
    \nonumber\\
    Y_d&=&i\sum_{\lambda\in\Delta_s}\left[
    g_{s_1}y_{d}(\lambda\cdot q,\xi)+
    g_{s_2}y_{d}^{(1/2)}(\lambda\cdot q, \xi)\right]E_{d}(\lambda),
    \nonumber\\
    E_{d}(\lambda)_{\mu \nu}&=&\delta_{\mu-\nu,2\lambda}.
    \label{eq:bnshXYrdef}
\end{eqnarray}
Both full and half-period  functions appear in these
matrix elements.  The diagonal parts of \(L_s\) and
\(M_s\) are given by
\begin{eqnarray}
    H_{\mu \nu}&=&\mu \delta_{\mu, \nu},\quad
    D_{\mu \nu}= \delta_{\mu, \nu}D_{\mu},
    \nonumber\\
    D_{\mu}&=&-i\left(g_{s_1}z_{d}(\mu\cdot q,\xi)+
    g_{s_2}z_{d}^{(1/2)}(\mu\cdot q, \xi)
    +g_l\sum_{\gamma\in\Delta_l,\
    \gamma\cdot\mu=1}z(\gamma\cdot q)\right).
    \label{eq:bnshHD}
\end{eqnarray}
The  functions \(z^{(1/2)}\) and
\(x_d^{(1/2)},y_d^{(1/2)}={x_d^{(1/2)}}^\prime\)  are the
same as those that appear in the Lax pairs for the twisted
\(C_n\) models. It is easy to verify
\begin{equation}
   Tr(L_s^2)=4{\cal H}_{B_n}^{ex-twisted},
   \label{bnshl2}
\end{equation}
in which the extended twisted \(B_n\) Hamiltonian is given above
(\ref{exbnHam}).

\subsubsection{Root type Lax pair for extended twisted \(B_n\)
model based on long roots
\(\Delta_l\)}
\label{bnshort}
The Lax pair is given in terms of the long roots.
The matrix elements of \(L_l\) and \(M_l\)  are
labeled  by indices
\(\alpha,\beta\) etc.
\noindent From the root difference pattern of
the long roots
\begin{equation}
   B_n:\qquad \mbox{long root}
   - \mbox{long root}=\left\{
   \begin{array}{l}
      \mbox{long root}\\
      2\times \mbox{long root}\\
      2\times \mbox{short root}\\
      \mbox{non-root}
   \end{array}
   \right.
   \label{bnlonglong}
\end{equation}
we obtain the Lax pair:
\begin{eqnarray}
    L_l(q,p,\xi) & = & p\cdot H + X + X_{d}+X_s,
   \quad
    \nonumber\\
    M_l(q,\xi) & = & D+Ds+Y+Y_{d}+Y_s.
    \label{eq:bnlnLaxform}
\end{eqnarray}
Here \(X\) and \(Y\) correspond to the part
of ``long root $-$ long root $=$ long root'' of
(\ref{bnlonglong}):
\begin{equation}
    X=ig_l\sum_{\alpha\in\Delta_l}x(\alpha\cdot
    q, \xi)E(\alpha),\quad
    Y=ig_l\sum_{\alpha\in\Delta_l}y(\alpha\cdot
    q, \xi)E(\alpha),\quad
    E(\alpha)_{\beta \gamma}=\delta_{\beta-\gamma,\alpha},
    \label{eq:bnlnXYdef}
\end{equation}
and $X_d$ and $Y_d$ correspond to
``long root $-$ long root $=$ 2\(\times\) long root''
of (\ref{bnlonglong}):
\begin{equation}
    X_d=2ig_l\sum_{\alpha\in\Delta_l}
    x_{d}(\alpha\cdot q, \xi)E_{d}(\alpha),\quad
    Y_d=ig_l\sum_{\alpha\in\Delta_l}
    y_{d}(\alpha\cdot q, \xi)E_{d}(\alpha),\quad
    E_{d}(\alpha)_{\beta \gamma}
    =\delta_{\beta-\gamma,2\alpha}.
    \label{eq:bnlnXYrdef}
\end{equation}
An additional term in \(L_l\) (\(M_l\)),
$X_s$ ($Y_s$) corresponds to
``long root $-$ long root $=$ 2\(\times\) short root''
of (\ref{bnlonglong}):
\begin{eqnarray}
    X_s&=&2i\sum_{\lambda\in\Delta_s}\left[
    g_{s_1}x_{d}(\lambda\cdot q,\xi)
    +g_{s_2}x_{d}^{(1/2)}(\lambda\cdot q,\xi)
    \right]E_{d}(\lambda),\nonumber\\
    Y_s&=&i\sum_{\lambda\in\Delta_s}\left[
    g_{s_1}y_{d}(\lambda\cdot q,\xi)
    +g_{s_2}y_{d}^{(1/2)}(\lambda\cdot q,\xi)
    \right]E_{d}(\lambda),\nonumber\\
    E_{d}(\lambda)_{\beta \gamma}
    &=&\delta_{\beta-\gamma,2\lambda}.
    \label{eq:bnlnXYsdef}
\end{eqnarray}
The diagonal parts of \(L_l\) and \(M_l\) are given by
\begin{equation}
    H_{\beta \gamma}=\beta \delta_{\beta, \gamma},\quad
    D_{\beta \gamma}= \delta_{\beta, \gamma}D_{\beta},\quad
    D_{\beta}=-ig_l\left(z(\beta\cdot
    q)+\sum_{\kappa\in\Delta_l,\
    \kappa\cdot\beta=1}z(\kappa\cdot q)\right),
   \label{eq:bnlnHD}
\end{equation}
and
\begin{equation}
    (Ds)_{\beta \gamma}= \delta_{\beta,
\gamma}(Ds)_{\beta},\quad
    (Ds)_{\beta}=-i\sum_{\lambda\in\Delta_s,\
    \beta\cdot\lambda=1}\left[
    g_{s_1}z_{d}(\lambda\cdot q,\xi)
    +g_{s_2}z_{d}^{(1/2)}(\lambda\cdot q,\xi)
    \right].
    \label{eq:bnlndsHD}
\end{equation}
The  functions \(z^{(1/2)}\) and
\(x_d^{(1/2)},y_d^{(1/2)}={x_d^{(1/2)}}^\prime\)  are the
same as those that appear in the Lax pairs for the twisted
\(C_n\) models. It is easy to verify that
\begin{equation}
   Tr(L_l^2)=8(n-1){\cal H}_{B_n}^{ex-twisted},
   \label{bnlnl2}
\end{equation}
in which the extended twisted \(B_n\) Hamiltonian is  given
above (\ref{exbnHam}).

\subsection{Twisted \(F_4\) model}
The set of \(F_4\) roots consists of two parts,
long  and short roots:
\begin{equation}
   \Delta_{F_4}=\Delta_l\cup\Delta_s,
\end{equation}
in which the roots are conveniently expressed in terms of
an orthonormal basis of \({\bf R}^4\):
\begin{eqnarray}
   \Delta_l&=&\{\alpha,\beta,\gamma,\ldots,\}=\{\pm e_j\pm
   e_k:
   \quad j\neq k=1,\ldots,4\},
   \quad 24\ \mbox{roots},
   \nonumber\\
   \Delta_s&=&\{\lambda,\mu,\nu,\ldots, \}=\{\pm e_j,
   {1\over2}(\pm e_1\pm e_2\pm e_3\pm e_4):
   \ j=1,\ldots,4\},
   \quad 24\ \mbox{roots.}
   \label{f4roots}
\end{eqnarray}
The set of long roots has the same structure as the \(D_4\)
roots and the set of short roots has the same structure as
the union of \(D_4\) vector, spinor and anti-spinor weights.
The untwisted \(F_4\) Hamiltonian is given by
\begin{equation}
    {\cal H}_{F_4}^{untwisted}={1\over2}p^2+{g_l^2\over2}
   \sum_{\alpha\in\Delta_l}
    \wp(\alpha\cdot q)+
   {g_s^2}\sum_{\lambda\in\Delta_s}
    \wp(\lambda\cdot q).
    \label{eq:untwistf4genham}
\end{equation}
It should be noted that  this has the same general structure
as the Hamiltonian of the untwisted \(B_n\) theory
(\ref{eq:untwistbngenham}).
As in the \(C_n\) and \(B_n\) cases, the {\em twisted} model is
obtained by multiplying one of the primitive periods of the
potential for the short roots by a factor of one half (II.3.41):
\begin{equation}
    {\cal H}_{F_4}^{twisted}={1\over2}p^2+{g_l^2\over2}
   \sum_{\alpha\in\Delta_l}
    \wp(\alpha\cdot q)+
   {g_s^2}\sum_{\lambda\in\Delta_s}
    \wp^{(1/2)}(\lambda\cdot q).
    \label{eq:twistf4genham}
\end{equation}

\subsubsection{Root type Lax pair for twisted \(F_4\)
model based on short roots
\(\Delta_s\)}
\label{f4short}
The Lax pair is given in terms of the short roots. %
It is
determined by  the root difference pattern of
the short roots:
\begin{equation}
   F_4:\qquad \mbox{short root}
   - \mbox{short root}=\left\{
   \begin{array}{l}
      \mbox{long root}\\
       \mbox{short root}\\
      2\times \mbox{short root}\\
      \mbox{non-root}
   \end{array}
   \right.
   \label{f4shrtshrt}
\end{equation}
The matrix elements of \(L_s\) and \(M_s\)  are
labeled  by indices
\(\mu,\nu\) etc.:
\begin{eqnarray}
    L_s(q,p,\xi) & = & p\cdot H + X + X_{d}+X_l, \nonumber\\
    M_s(q,\xi) & = & D+D_l+Y+Y_{d}+Y_l.
    \label{eq:f4shLaxform}
\end{eqnarray}
Here \(X\) and \(Y\) correspond to the part
of ``short root $-$ short root $=$ short root'' of
(\ref{f4shrtshrt}):
\begin{equation}
    X=ig_s\sum_{\lambda\in\Delta_s}x^{(1/2)}(\lambda\cdot
    q, \xi)E(\lambda),\quad
    Y=ig_s\sum_{\lambda\in\Delta_s}y^{(1/2)}(\lambda\cdot
    q, \xi)E(\lambda),\quad
    E(\lambda)_{\mu \nu}=\delta_{\mu-\nu,\lambda},
    \label{eq:f4shXYdef}
\end{equation}
and $X_d$ and $Y_d$ correspond to
``short root $-$ short root $=$ 2\(\times\) short
root'' of (\ref{f4shrtshrt}):
\begin{equation}
    X_d=2ig_s\sum_{\lambda\in\Delta_s}
    x_{d}^{(1/2)}(\lambda\cdot q, \xi)E_{d}(\lambda),\quad
    Y_d=ig_s\sum_{\lambda\in\Delta_s}
    y_{d}^{(1/2)}(\lambda\cdot q, \xi)E_{d}(\lambda),\quad
    E_{d}(\lambda)_{\mu \nu}=\delta_{\mu-\nu,2\lambda},
    \label{eq:f4shXYrdef}
\end{equation}
and these functions have half-period.
The additional terms \(X_l\) and \(Y_l\) correspond to
``short root $-$ short root $=$  long
root'' of (\ref{f4shrtshrt}) and they have the same form as in the
untwisted model:
\begin{equation}
    X_l=ig_l\sum_{\alpha\in\Delta_l}x(\alpha\cdot
    q, \xi)E(\alpha),\quad
    Y_l=ig_l\sum_{\alpha\in\Delta_l}y(\alpha\cdot
    q, \xi)E(\alpha),\quad
    E(\alpha)_{\mu \nu}=\delta_{\mu-\nu,\alpha}.
    \label{eq:f4shXYlndef}
\end{equation}
The diagonal parts of \(L_s\) and \(M_s\) are given by
\begin{equation}
    H_{\mu \nu}=\mu \delta_{\mu, \nu},\quad
    D_{\mu \nu}= \delta_{\mu, \nu}D_{\mu},\quad
    D_{\mu}=-ig_s\left(\,z^{(1/2)}(\mu\cdot
   q)+\sum_{\lambda\in\Delta_s,\
    \lambda\cdot\mu=1/2}z^{(1/2)}(\lambda\cdot q)\right),
    \label{eq:f4shHD}
\end{equation}
and
\begin{equation}
   (D_l)_{\mu \nu}= \delta_{\mu,
   \nu}(D_l)_{\mu},\quad
    (D_l)_{\mu}=-ig_l\sum_{\alpha\in\Delta_l,\
    \alpha\cdot\mu=1}z(\alpha\cdot q).
    \label{eq:f4shdLHD}
\end{equation}
The  functions \(x^{(1/2)},y^{(1/2)},z^{(1/2)}\) and
\(x_d^{(1/2)},y_d^{(1/2)}\) are the same as those appear in
the Lax pairs for the twisted \(C_n\) models. It is easy to
verify that
\begin{equation}
   Tr(L_s^2)=12{\cal H}_{F_4}^{twisted},
   \label{f4shl2}
\end{equation}
in which the twisted Hamiltonian is given by
(\ref{eq:twistf4genham}). In different notation
the above Lax pair (\ref{eq:f4shLaxform})--(\ref{eq:f4shdLHD}) was reported
in
\cite{DHPh}.

\subsubsection{Root type Lax pair for untwisted \(F_4\)
model based on long roots
\(\Delta_l\)}
\label{f4long}
The Lax pair is given in terms of long roots. 
The general structure of this Lax pair is essentially the same as
that of the \(B_n\) theory, since the pattern of the long root--
long root
\begin{equation}
   F_4:\qquad \mbox{long root}
   - \mbox{long root}=\left\{
   \begin{array}{l}
      \mbox{long root}\\
      2\times \mbox{long root}\\
      2\times \mbox{short root}\\
      \mbox{non-root}
   \end{array}
   \right.
   \label{f4longlong}
\end{equation}
is the same as that
of \(B_n\) (\ref{bnlonglong}).
So we list the general form only without further explanation.
They are
matrices with indices
\(\beta,\gamma\) etc.:
\begin{eqnarray}
    L_l(q,p,\xi) & = & p\cdot H + X + X_{d}+X_s, \nonumber\\
    M_l(q,\xi) & = & D+Ds+Y+Y_{d}+Y_s.
    \label{eq:f4lnLaxform}
\end{eqnarray}
\begin{equation}
    X=ig_l\sum_{\alpha\in\Delta_l}x(\alpha\cdot
    q, \xi)E(\alpha),\quad
    Y=ig_l\sum_{\alpha\in\Delta_l}y(\alpha\cdot
    q, \xi)E(\alpha),\quad
    E(\alpha)_{\beta \gamma}=\delta_{\beta-\gamma,\alpha}.
    \label{eq:f4lnXYdef}
\end{equation}
\begin{equation}
    X_d=2ig_l\sum_{\alpha\in\Delta_l}
    x_{d}(\alpha\cdot q, \xi)E_{d}(\alpha),\quad
    Y_d=ig_l\sum_{\alpha\in\Delta_l}
    y_{d}(\alpha\cdot q, \xi)E_{d}(\alpha),\quad
    E_{d}(\alpha)_{\beta \gamma}
    =\delta_{\beta-\gamma,2\alpha}.
    \label{eq:f4lnXYrdef}
\end{equation}
\begin{equation}
   X_s=2ig_s\sum_{\lambda\in\Delta_s}
   x_{d}^{(1/2)}(\lambda\cdot q, \xi)E_{d}(\lambda),\quad
   Y_s=ig_s\sum_{\lambda\in\Delta_s}
   y_{d}^{(1/2)}(\lambda\cdot q, \xi)E_{d}(\lambda),\quad
   E_{d}(\lambda)_{\beta \gamma}
   =\delta_{\beta-\gamma,2\lambda}.
   \label{eq:f4lnXYsdef}
\end{equation}
The diagonal parts of \(L_l\) and \(M_l\) are given by
\begin{equation}
    H_{\beta \gamma}=\beta \delta_{\beta, \gamma},\quad
    D_{\beta \gamma}= \delta_{\beta, \gamma}D_{\beta},\quad
    D_{\beta}=-ig_l\left(z(\beta\cdot q)
    +\sum_{\kappa\in\Delta_l,\
    \kappa\cdot\beta=1}z(\kappa\cdot q)\right),
   \label{eq:f4lnHD}
\end{equation}
and
\begin{equation}
    (Ds)_{\beta \gamma}= \delta_{\beta,
\gamma}(Ds)_{\beta},\quad
    (Ds)_{\beta}=-ig_s\sum_{\lambda\in\Delta_s,\
    \beta\cdot\lambda=1}z^{(1/2)}(\lambda\cdot q).
    \label{eq:f4lndsHD}
\end{equation}
The  functions \(z^{(1/2)}\) and
\(x_d^{(1/2)},y_d^{(1/2)}\) are the same as those that appear in
the Lax pairs for the twisted \(C_n\) models.
It is easy to verify that
\begin{equation}
   Tr(L_l^2)=24{\cal H}_{F_4}^{twisted},
   \label{f4lnl2}
\end{equation}
in which the twisted \(F_4\) Hamiltonian is  given above
(\ref{eq:twistf4genham}).

\subsection{Twisted \(G_2\) model}
A Lax pair for the twisted \(G_2\) model has never been
constructed in any form. Here we present new Lax pairs
based on short and long roots for this model.
The twisted
\(G_2\) model has some  features different from those
discussed above due to the ratio of (long
root)\({}^2\)/(short root)\({}^2=3\) instead of 2 for the
other non-simply laced root systems. The set of
\(G_2\) roots consists of two parts,  long  and short roots:
\begin{equation}
   \Delta_{G_2}=\Delta_l\cup\Delta_s,
\end{equation}
in which the roots are conveniently expressed in terms of
an orthonormal basis of \({\bf R}^2\):
\begin{eqnarray}
   \Delta_l&=&\{\alpha,\beta,\gamma,\ldots,\}=
   \{\pm(-3e_1+\sqrt3e_2)/2,
   \pm(3e_1+\sqrt3e_2)/2,
   \pm\sqrt3 e_2\},
   \quad 6\ \mbox{roots},
   \nonumber\\
   \Delta_s&=&\{\lambda,\mu,\nu,\ldots, \}=\{\pm e_1,
   \pm(-e_1+\sqrt3e_2)/2,\pm(e_1+\sqrt3e_2)/2\},
   \quad 6\ \mbox{roots.}
   \label{g2roots}
\end{eqnarray}
The sets of long and short roots have the same structure as the
\(A_2\) roots, scaled
and rotated by
\({\pi/6}\).
The untwisted \(G_2\) Hamiltonian is given by
\begin{equation}
    {\cal H}_{G_2}^{untwisted}={1\over2}p^2+{g_l^2\over3}
   \sum_{\alpha\in\Delta_l}
    \wp(\alpha\cdot q)+
   {g_s^2}\sum_{\lambda\in\Delta_s}
    \wp(\lambda\cdot q).
    \label{eq:untwistg2genham}
\end{equation}
The twisting leads to  a short root potential with a
third-period function (II.3.48):
\begin{equation}
    {\cal H}_{G_2}^{twisted}={1\over2}p^2+{g_l^2\over3}
    \sum_{\alpha\in\Delta_l}
    \wp(\alpha\cdot q)+
   {g_s^2}\sum_{\lambda\in\Delta_s}
    \wp^{(1/3)}(\lambda\cdot q),
    \label{eq:twistg2genham}
\end{equation}
in which we choose the convention (\ref{halfperiods})
\begin{eqnarray}
   \wp^{(1/3)}(u)&\equiv&\wp(u|
   2\omega_1/3,2\omega_3)\nonumber\\
   &=&
   \wp(u)+\wp(u+2\omega_1/3)+\wp(u+4\omega_1/3)-
   \wp(2\omega_1/3)-\wp(4\omega_1/3).
   \label{thirdperiods}
\end{eqnarray}
The second equality is Landen's transformation for a third-period.

\subsubsection{Root type Lax pair for twisted \(G_2\)
model based on short roots
\(\Delta_s\)}
\label{g2short}
The Lax pair is given in terms of short roots. 
The general structure of this Lax pair is essentially the same as
that of \(F_4\) theory, since the pattern of the short root--
short root
\begin{equation}
   G_2:\qquad \mbox{short root}
   - \mbox{short root}=\left\{
   \begin{array}{l}
      \mbox{long root}\\
       \mbox{short root}\\
      2\times \mbox{short root}\\
      \mbox{non-root}
   \end{array}
   \right.
   \label{g2shrtshrt}
\end{equation}
 is the same as that of
the \(F_4\) model (\ref{f4shrtshrt}). The difference is that the
functions associated with the short root potential have
a third-period. The matrix
elements of
\(L_s\) and
\(M_s\)  are labeled  by indices
\(\mu,\nu\) etc.:
\begin{eqnarray}
    L_s(q,p,\xi) & = & p\cdot H + X + X_{d}+X_l, \nonumber\\
    M_s(q,\xi) & = & D+D_l+Y+Y_{d}+Y_l,
    \label{eq:g2shLaxform}
\end{eqnarray}
\begin{equation}
    X=ig_s\sum_{\lambda\in\Delta_s}x^{(1/3)}(\lambda\cdot
   q, \xi)E(\lambda),\quad
    Y=ig_s\sum_{\lambda\in\Delta_s}y^{(1/3)}(\lambda\cdot
   q, \xi)E(\lambda),\quad
    E(\lambda)_{\mu \nu}=\delta_{\mu-\nu,\lambda}.
    \label{eq:g2shXYdef}
\end{equation}
\begin{equation}
    X_d=2ig_s\sum_{\lambda\in\Delta_s}
   x_{d}^{(1/3)}(\lambda\cdot q, \xi)E_{d}(\lambda),\quad
    Y_d=ig_s\sum_{\lambda\in\Delta_s}
   y_{d}^{(1/3)}(\lambda\cdot q, \xi)E_{d}(\lambda),\quad
    E_{d}(\lambda)_{\mu \nu}=\delta_{\mu-\nu,2\lambda}.
    \label{eq:g2shXYrdef}
\end{equation}
\begin{equation}
    X_l=ig_l\sum_{\alpha\in\Delta_l}x(\alpha\cdot
   q, \xi)E(\alpha),\quad
    Y_l=ig_l\sum_{\alpha\in\Delta_l}y(\alpha\cdot
   q, \xi)E(\alpha),\quad
    E(\alpha)_{\mu \nu}=\delta_{\mu-\nu,\alpha}.
    \label{eq:g2shXYlndef}
\end{equation}
The diagonal parts of \(L_s\) and \(M_s\) are given by
\begin{equation}
    H_{\mu \nu}=\mu \delta_{\mu, \nu},\quad
    D_{\mu \nu}= \delta_{\mu, \nu}D_{\mu},\quad
    D_{\mu}=-ig_s\left(\,z^{(1/3)}(\mu\cdot
    q)+\sum_{\lambda\in\Delta_s,\
    \lambda\cdot\mu=1/2}z^{(1/3)}(\lambda\cdot q)\right),
    \label{eq:g2shHD}
\end{equation}
and
\begin{equation}
   (D_l)_{\mu \nu}= \delta_{\mu,
   \nu}(D_l)_{\mu},\quad
    (D_l)_{\mu}=-ig_l\sum_{\alpha\in\Delta_l,\
    \alpha\cdot\mu=3/2}z(\alpha\cdot q).
    \label{eq:g2shdLHD}
\end{equation}
The new functions \(x^{(1/3)},y^{(1/3)},z^{(1/3)}\) and
\(x_d^{(1/3)},y_d^{(1/3)}\) must satisfy certain sum rules
which are generalisations of the third sum rule.
These sum rules are listed in section~\ref{twfuns} together
with the explicit forms of the new functions.
 It is easy to verify that
\begin{equation}
   Tr(L_s^2)=6{\cal H}_{G_2}^{twisted},
   \label{g2shl2}
\end{equation}
in which the twisted \(G_2\) Hamiltonian is given by
(\ref{eq:twistg2genham}).

\subsubsection{Root type Lax pair for untwisted \(G_2\)
model based on long roots
\(\Delta_l\)}
\label{g2long}
This Lax pair is different from the others because of the
`triple root' term in the pattern of the long root--
long root:
\begin{equation}
   G_2:\qquad \mbox{long root}
   - \mbox{long root}=\left\{
   \begin{array}{l}
      \mbox{long root}\\
      2\times \mbox{long root}\\
      3\times \mbox{short root}\\
      \mbox{non-root}
   \end{array}
   \right.
   \label{g2longlong}
\end{equation}
The matrix elements of \(L_l\) and \(M_l\)  are
labeled  by indices
\(\beta,\gamma\) etc.:
\begin{eqnarray}
    L_l(q,p,\xi) & = & p\cdot H + X + X_{d}+X_t, \nonumber\\
    M_l(q,\xi) & = & D+Dt+Y+Y_{d}+Y_t.
    \label{eq:g2lnLaxform}
\end{eqnarray}
The terms \(X,Y\) and \(X_d,Y_d\) are the same as before:
\begin{equation}
    X=ig_l\sum_{\alpha\in\Delta_l}x(\alpha\cdot
   q, \xi)E(\alpha),\quad
    Y=ig_l\sum_{\alpha\in\Delta_l}y(\alpha\cdot
   q, \xi)E(\alpha),\quad
    E(\alpha)_{\beta \gamma}=\delta_{\beta-\gamma,\alpha}.
    \label{eq:g2lnXYdef}
\end{equation}
\begin{equation}
    X_d=2ig_l\sum_{\alpha\in\Delta_l}
   x_{d}(\alpha\cdot q, \xi)E_{d}(\alpha),\quad
    Y_d=ig_l\sum_{\alpha\in\Delta_l}
   y_{d}(\alpha\cdot q, \xi)E_{d}(\alpha),\quad
    E_{d}(\alpha)_{\beta \gamma}
   =\delta_{\beta-\gamma,2\alpha}.
    \label{eq:g2lnXYrdef}
\end{equation}
The `triple root' terms \(X_t\) and \(Y_t\) are associated with the
 new functions \(x_t^{(1/3)}\) and
\(y_t^{(1/3)}\):
\begin{equation}
    X_t=3ig_s\sum_{\lambda\in\Delta_s}
   x_{t}^{(1/3)}(\lambda\cdot q, \xi)E_{t}(\lambda),\quad
    Y_t=ig_s\sum_{\lambda\in\Delta_s}
   y_{t}^{(1/3)}(\lambda\cdot q, \xi)E_{t}(\lambda),\quad
    E_{t}(\lambda)_{\beta \gamma}
   =\delta_{\beta-\gamma,3\lambda}.
    \label{eq:g2lnXYtdef}
\end{equation}
The diagonal parts of \(L_l\) and \(M_l\) are given by
\begin{equation}
    H_{\beta \gamma}=\beta \delta_{\beta, \gamma},\quad
    D_{\beta \gamma}= \delta_{\beta, \gamma}D_{\beta},\quad
    D_{\beta}=-ig_l\left(z(\beta\cdot q)
   +\sum_{\kappa\in\Delta_l,\
    \kappa\cdot\beta=3/2}z(\kappa\cdot q)\right),
   \label{eq:g2lnHD}
\end{equation}
and
\begin{equation}
    (Dt)_{\beta \gamma}= \delta_{\beta,
\gamma}(Dt)_{\beta},\quad
    (Dt)_{\beta}=-ig_s\sum_{\lambda\in\Delta_s,\
    \beta\cdot\lambda=3/2}z^{(1/3)}(\lambda\cdot q).
    \label{eq:g2lndtHD}
\end{equation}
The new functions
\(x_t^{(1/3)}\) and \(y_t^{(1/3)}\) must satisfy certain sum
rules which are generalisations of the third sum rule.
These sum rules are listed in section~\ref{twfuns} together
with the explicit forms of the new functions.

\subsection{Extended Twisted \(BC_n\) root system Lax
pair with five independent couplings}
The \(BC_n\) root system consists of three parts:
long,  middle and short roots:
\begin{equation}
   \Delta_{BC_n}=\Delta_l\cup\Delta_m\cup\Delta_s,
\end{equation}
in which the roots are conveniently expressed in terms of
an orthonormal basis of \({\bf R}^n\):
\begin{eqnarray}
   \Delta_l&=&\{\Xi,\Upsilon,\Omega,\ldots,\}=\{\pm 2e_j:
   \qquad \quad j=1,\ldots,n\},
   \quad 2n\ \mbox{roots},
   \label{bcnLnroots}\\
   \Delta_m&=&\{\alpha,\beta,\gamma,\ldots, \}=
   \{\pm e_j\pm e_k: \quad j,k=1,\ldots,n\},\quad
   2n(n-1)\ \mbox{roots},
   \label{bcnmdroots}\\
   \Delta_s&=&\{\lambda,\mu,\nu,\ldots, \}
   =\{\pm e_j: \qquad \quad
   j=1,\ldots,n\},
   \quad \quad \ 2n\ \mbox{roots.}
   \label{bcnroots}
\end{eqnarray}
The Hamiltonian of the untwisted  \(BC_n\) model is the \(C_n\)
Hamiltonian  (\ref{eq:untwistcngenham}) plus
the contribution from the short root potential:
\begin{equation}
    {\cal H}_{BC_n}^{untwisted}={1\over2}p^2+{g_m^2\over2}
    \sum_{\alpha\in\Delta_m}
    \wp(\alpha\cdot q)+{g_l^2\over4}\sum_{\Xi\in\Delta_l}
    \wp(\Xi\cdot q)+{g_s^2}\sum_{\lambda\in\Delta_s}
    \wp(\lambda\cdot q).
    \label{eq:untwistedbcngenham}
\end{equation}
The Hamiltonian of the extended twisted  \(BC_n\) model is,
roughly speaking,
the sum of the extended twisted
\(C_n\) Hamiltonian (\ref{excnHam}) and  the extended twisted
\(B_n\) Hamiltonian (\ref{exbnHam}) at half-period:
\begin{eqnarray}
    {\cal H}_{BC_n}^{ex-twisted}&=&{1\over2}p^2+{g_m^2\over2}
    \sum_{\alpha\in\Delta_m}
    \wp^{(1/2)}(\alpha\cdot q)+{1\over4}\sum_{\Xi\in\Delta_l}
    \left[\tilde{g}^2_{l_1}\wp(\Xi\cdot q)
     +g^2_{l_2}\wp^{(1/2)}(\Xi\cdot
     q)\right]\nonumber\\
    &&+ \sum_{\lambda\in\Delta_s}
    \left[\tilde{g}^2_{s_1}\wp^{(1/2)}(\lambda\cdot q)+
     \tilde{g}^2_{s_2}\wp^{(1/4)}(\lambda\cdot q)\right],
    \label{extwistbcnham}
\end{eqnarray}
in which renormalised coupling constants are defined by
\begin{eqnarray}
   \tilde{g}^2_{l_1}&=&g_{l_1}(g_{l_1}+2g_{l_2}),\nonumber\\
   \tilde{g}^2_{s_1}&=&g_{s_1}\left(g_{s_1}
   +{1\over2}(g_{l_1}+g_{l_2})
   +2g_{s_2}\right)+{1\over2}g_{l_1}g_{s_2},\nonumber\\
   \tilde{g}^2_{s_2}&=&g_{s_2}(g_{s_2}+2g_{l_2}).
   \label{threeren}
\end{eqnarray}
The twisted \(BC_n\) Hamiltonian derived by reduction in paper II
(II.3.61) may be obtained by
\[
   g_{l_1}=g_{m}=g_{s_1}=-g_{s_2}=\sqrt2g,\quad g_{l_2}=0.
\]

Here we consider the Lax pair based on the middle roots only.
The pattern of middle root-- middle root  is
\begin{equation}
   BC_n:\qquad \mbox{middle root}
   - \mbox{middle root}=\left\{
   \begin{array}{l}
      \mbox{long root}\\
      \mbox{middle root}\\
      2\times \mbox{middle root}\\
      2\times \mbox{short root}\\
      \mbox{non-root}
   \end{array}
   \right.
   \label{bcnmdmd}
\end{equation}
We can then construct the root type Lax
pair for \(BC_n\) root system:
\begin{eqnarray}
    L_m(q,p,\xi) & = & p\cdot H + X + X_{d}+X_L+X_s, \nonumber\\
    M_m(q,\xi) & = & D+D_L+Y+Y_{d}+Y_L+Ds+Y_s.
    \label{eq:bcnmdLaxform}
\end{eqnarray}
The matrix elements of \(L_m\) and \(M_m\)  are
labeled  by indices
\(\beta,\gamma\) etc. This Lax pair is, roughly speaking,
the sum of the extended twisted \(C_n\) Lax pair based on short
roots (with renaming of the couplings) and the extended twisted
\(B_n\) Lax pair based on long roots (both periods of the
potentials are halved). Here
\(p\cdot H + X + X_{d}+X_L\) (\(D+D_L+Y+Y_{d}+Y_L\))  is exactly the
same as the
\(L_s\) (\(M_s\)) matrix of extended \(C_n\) models with three coupling
constants based on short roots.
An additional term in \(L_m\) (\(M_m\)),
$X_s$ ($Y_s$) corresponds to
``middle root $-$ middle root $=$ 2\(\times\) short root''
of (\ref{bcnmdmd}):
\begin{eqnarray}
    X_s&=&2i\sum_{\lambda\in\Delta_s}\left[
    g_{s_1}x_{d}^{(1/2)}(\lambda\cdot q,\xi)+
    g_{s_2}x_{d}^{(1/4)}(\lambda\cdot q,
    \xi)\right]E_{d}(\lambda),
    \nonumber\\
    Y_s&=&i\sum_{\lambda\in\Delta_s}\left[
    g_{s_1}y_{d}^{(1/2)}(\lambda\cdot q,\xi)+
    g_{s_2}y_{d}^{(1/4)}(\lambda\cdot q,
\xi)\right]E_{d}(\lambda),
\nonumber\\
    E_{d}(\lambda)_{\mu \nu}&=&\delta_{\mu-\nu,2\lambda},
    \label{eq:bncnshXYrdef}
\end{eqnarray}
\begin{equation}
    (Ds)_{\beta \gamma}= \delta_{\beta,
\gamma}(Ds)_{\beta},\quad
    (Ds)_{\beta}=-i\sum_{\lambda\in\Delta_s,\
    \beta\cdot\lambda=1}\left[
    g_{s_1}z_{d}^{(1/2)}(\lambda\cdot q,\xi)+
    g_{s_2}z_{d}^{(1/4)}(\lambda\cdot q, \xi)\right],
    \label{eq:bcnlndsHD}
\end{equation}
which is a half-period version of  the
extended twisted \(B_n\) model based on long roots
(\ref{eq:bnlnXYsdef}).

The new functions
\(x_d^{(1/4)},y_d^{(1/4)},z_d^{(1/4)}\) and their sum
rules are  given in section 5. It is easy to
verify that
\begin{equation}
   Tr(L_m^2)=8(n-1){\cal H}_{BC_n}^{ex-twisted},
   \label{bcnshl2}
\end{equation}
for which various factorisation relations
(\ref{secfac}),(\ref{fac1d1})--(\ref{fac-14d}), are responsible
for the renormalisation of the couplings (\ref{threeren}).
The verification
that the Lax equation is equivalent to the canonical
equation of motion is again essentially the same as in the extended
twisted \(C_n\) and \(B_n\) models.

At the end of this section let us remark on the relation between our
extended twisted \(BC_n\) model and Inozemtsev's work \cite{Ino}.
His model has a potential, which  in our notation is:
\begin{eqnarray}
    V_{Inozemtsev}&=&
    {g^2\over2}
    \sum_{\alpha\in\Delta_m}
    \wp^{(1/2)}(\alpha\cdot q)\nonumber\\
    &&+ {1\over2}\sum_{\lambda\in\Delta_s}
    \left[g^2_{1}\wp^{(1/2)}(\lambda\cdot q)+
     g^2_{2}\wp^{(1/2)}(\lambda\cdot q+{\omega_1\over2})+
     g^2_{3}\wp^{(1/2)}(\lambda\cdot q+\omega_3)\right.
    \nonumber\\
    &&\left.\qquad \qquad \quad +\
     g^2_{4}\wp^{(1/2)}(\lambda\cdot q
  +{\omega_1\over2}+\omega_3)\right],
    \label{inopoten}
\end{eqnarray}
which is equivalent to our Hamiltonian (\ref{extwistbcnham})
if the coupling constants are related by
\begin{eqnarray}
     \tilde{g}^2_{l_1}&=&8(g_{3}^2-g_{4}^2),\quad
      g^2_{l_2}=8g_{4}^2,\quad
      g^2_{m}=g^2,\nonumber\\
     \tilde{g}^2_{s_1}&=&{1\over2}(g_{1}^2-g_{2}^2-
     g_{3}^2+g_{4}^2),\quad
     \tilde{g}^2_{s_2}={1\over2}(g_{2}^2-g_{4}^2).
     \label{ino-our-coupl}
\end{eqnarray}

\section{Functions in the Lax pairs for twisted models}
\label{twfuns}
\setcounter{equation}{0}
As we have seen in the previous section, many different
functions enter in the Lax pairs for the twisted
Calogero-Moser models.
They are untwisted functions
\[
   x,\quad y=x^\prime,\quad z,\qquad x_d,\quad
   y_d=x_d^\prime,\quad z_d
\]
associated with the potential for the long roots.
Another set of functions with half-period
\[
   x^{(1/2)},\quad y^{(1/2)}={x^{(1/2)}}^\prime,\quad
   z^{(1/2)},\qquad x_d^{(1/2)},\quad
   y_d^{(1/2)}={x_d^{(1/2)}}^\prime,\quad z_d^{(1/2)}
\]
is introduced for potential for short roots having the
ratio (long root)\({}^2:\)(short
root)\({}^2=2:1\).
They appear in the twisted \(C_n, B_n,F_4\) and \(BC_n\)
models. The third set of functions with third-period
\[
   x^{(1/3)},\quad y^{(1/3)}={x^{(1/3)}}^\prime,\quad
   z^{(1/3)},\qquad x_d^{(1/3)},\quad
   y_d^{(1/3)}={x_d^{(1/3)}}^\prime,\quad z_d^{(1/3)},
\]
\[
   x_t^{(1/3)},\quad y_t^{(1/3)}={x_t^{(1/3)}}^\prime,\quad
   z_t^{(1/3)}
\]
is used in the twisted \(G_2\) model
in which the ratio (long root)\({}^2:\)(short
root)\({}^2\) is \(3:1\).
Finally the functions with quarter period
\[
   x_d^{(1/4)},\quad y_d^{(1/4)}={x_d^{(1/4)}}^\prime,\quad
   z_d^{(1/4)}
\]
 appear in the twisted \(BC_n\) model.

\subsection{Twisted sum rules}
In this subsection we give the sum rules or the functional
relations to be satisfied by these functions.
The explicit forms of these functions will be given in the
next subsection. First
of all, they all have to satisfy the {\em first sum rule}:
\begin{equation}
   {x^{(1/n)}}^\prime(u)x^{(1/n)}(v)-{x^{(1/n)}}^\prime(v)
    x^{(1/n)}(u)=
    x^{(1/n)}(u+v)[\wp^{(1/n)}(v)-\wp^{(1/n)}(u)],\quad n=2,3,
   \label{eq:ident1n}
\end{equation}
\begin{equation}
    {x_d^{(1/n)}}^\prime(u)x_d^{(1/n)}(v)-{x_d^{(1/n)}}^\prime(v)
    x_d^{(1/n)}(u)=
    x_d^{(1/n)}(u+v)[\wp^{(1/n)}(v)-\wp^{(1/n)}(u)],\quad
    n=2,3,4,
    \label{eq:ident1nd}
\end{equation}
\begin{equation}
   {x_t^{(1/3)}}^\prime(u)x_t^{(1/3)}(v)-
   {x_t^{(1/3)}}^\prime(v)
   x_t^{(1/3)}(u)=
    x_t^{(1/3)}(u+v)[\wp^{(1/3)}(v)-\wp^{(1/3)}(u)],
    \label{eq:ident1nt}
\end{equation}
so that they give the potential with the \(1/n\)-th period
by the factorisation formula (\ref{eq:integ}):
\begin{eqnarray*}
   z^{(1/n)}(u)&=&x^{(1/n)}(u)x^{(1/n)}(-u)=
   -\wp^{(1/n)}(u)+const,\quad n=2,3,\\
   z_d^{(1/n)}(u)&=&x_d^{(1/n)}(u)x_d^{(1/n)}(-u)
   =-\wp^{(1/n)}(u)+const,\quad n=2,3,4,\\
   z_t^{(1/3)}(u)&=&x_t^{(1/3)}(u)x_t^{(1/3)}(-u)
   =-\wp^{(1/3)}(u)+const.
\end{eqnarray*}

There are three types of {\em twisted second sum
rules}:
\begin{eqnarray}
   x(u-v)\left[\wp^{(1/2)}(v)-\wp^{(1/2)}(u)\right]
   &+&2\left[x_d^{(1/2)}(u)\,y(-u-v)-
   y(u+v)\,x_d^{(1/2)}(-v)\right]\nonumber\\
   &&\hspace*{-2cm}
   +\ \ x(u+v)\,y_d^{(1/2)}(-v)-y_d^{(1/2)}(u)\,x(-u-v)=0,
    \label{eq:twsecsum1}
   \end{eqnarray}
   \begin{eqnarray}
   x^{(1/2)}(u-v)\left[\wp(2v)-\wp(2u)\right]
   &+&x(2u)\,y^{(1/2)}(-u-v)-
   y^{(1/2)}(u+v)\,x(-2v)\nonumber\\
   &&\hspace*{-2cm}
   +\ \ x^{(1/2)}(u+v)\,y(-2v)-y(2u)\,x^{(1/2)}(-u-v)=0,
    \label{eq:twsecsum2}
\end{eqnarray}
\begin{eqnarray}
   x_d^{(1/2)}(u-v)\left[\wp(2v)-\wp(2u)\right]
   &+&x_d(2u)\,y_d^{(1/2)}(-u-v)-
   y_d^{(1/2)}(u+v)\,x_d(-2v)\nonumber\\
   &&\hspace*{-2cm}
   +\ \ x_d^{(1/2)}(u+v)\,y_d(-2v)-y_d(2u)\,x_d^{(1/2)}(-u-v)=0,
    \label{eq:twsecsum3}
\end{eqnarray}
in which (\ref{eq:twsecsum1}) is obtained from the untwisted
second sum rule (\ref{eq:xydiden}) by changing the double
root functions \(x_d,y_d,z_d\) into
\(x_d^{(1/2)},y_d^{(1/2)},z_d^{(1/2)}\). This is related to
the two dimensional root system \(B_n\) and  a part of its
 root difference patterns
(\ref{bnshrtshrt}), (\ref{bnlonglong}), i.e.
\begin{equation}
  \mbox{(long or short) root}
   - \mbox{(long or short) root}=\left\{
   \begin{array}{l}
      \mbox{long root}\qquad \quad \ \ (x)\\
      2\times \mbox{short root} \quad (x_d^{(1/2)})\\
   \end{array}
   \right.
   \label{l-2short}
\end{equation}
This sum rule is responsible for the consistency
of the twisted Lax pairs of the models containing the above
pattern, see subsections \ref{cnshort}, \ref{bnshort},
\ref{bnlong}, \ref{f4short}, \ref{f4long}.
The next twisted second sum rule (\ref{eq:twsecsum2}) is
related to the two dimensional root system \(C_n\) and to a
part of its
 root difference patterns
(\ref{cnshrtshrt}), i.e.
\begin{equation}
   \mbox{short root}
   - \mbox{short root}=\left\{
   \begin{array}{l}
      \mbox{long root}\qquad \quad (x)\\
       \mbox{short root}\qquad (x^{(1/2)})\\
   \end{array}
   \right.
   \label{l-short}
\end{equation}
This sum rule is responsible for the consistency
of the twisted Lax pairs of the models containing the above
pattern, that is the \(C_n\) and \(F_4\) models based on
short roots, see subsections \ref{cnshort}, \ref{f4short}.
The last twisted second sum rule (\ref{eq:twsecsum3}) is
related to the two dimensional root system \(B_n (C_n)\) and
to a part of its
 root difference patterns (\ref{cnlonglong}),
(\ref{bnlonglong}), i.e.
\begin{equation}
  \mbox{long root}
   - \mbox{long root}=\left\{
   \begin{array}{l}
     2\times   \mbox{long root}\qquad \quad (x_d)\\
       2\times  \mbox{short root}\qquad (x_d^{(1/2)})\\
   \end{array}
   \right.
  \label{2l-2short}
\end{equation}
Obviously (\ref{eq:twsecsum3}) can be obtained from
(\ref{eq:twsecsum2}) by changing \(x,x^{(1/2)}\) (and
\(y,y^{(1/2)}\) ) to
\(x_d,x_d^{(1/2)}\) (and \(y_d,y_d^{(1/2)}\)). This sum rule is
responsible for the consistency of the twisted Lax pairs of
the models containing the above pattern, that is the
\(C_n\),
\(B_n\) and \(F_4\) models based on long roots, see subsections
\ref{cnlong},
\ref{bnlong}, \ref{f4long}.
Finally, the twisted functions also satisfy the ordinary
second sum rules:
\begin{eqnarray}
   &&\hspace*{-1cm}x^{(1/n)}(u-v)\left[\wp^{(1/n)}(v)-
   \wp^{(1/n)}(u)\right]
   +2\left[x_d^{(1/n)}(u)\,y^{(1/n)}(-u-v)-
   y^{(1/n)}(u+v)\,x_d^{(1/n)}(-v)\right]\nonumber\\
   &+&x^{(1/n)}(u+v)\,y_d^{(1/n)}(-v)-
   y_d^{(1/n)}(u)\,x^{(1/n)}(-u-v)=0,\qquad n=2,3,4.
    \label{eq:xydiden-n}
\end{eqnarray}
This is responsible for the consistency of the twisted Lax
pairs of the models containing the  pattern
\begin{equation}
   \mbox{short (middle) root}
   - \mbox{short (middle) root}=\left\{
   \begin{array}{l}
      \mbox{short (middle) root}\qquad \quad (x^{(1/n)})\\
    2\times   \mbox{short (middle) root}\qquad (x^{(1/n)}_d)\\
   \end{array}
   \right.
   \label{s-2short}
\end{equation}
This occurs in \(F_4\) (\(n=2\)) and \(G_2\) (\(n=3\)) Lax pair
based on short roots and \(BC_n\) (\(n=2\)) Lax pair
based on middle roots. Some of the above twisted second sum
rules, (\ref{eq:twsecsum1}), (\ref{eq:twsecsum2}) and
(\ref{eq:xydiden-n}) for \(n=2\) were reported in
\cite{DHPh}.

\bigskip
There are two types of {\em twisted third sum
rules}:
\begin{eqnarray}
   &&x(3u-3v)[\wp^{(1/3)}(2u-v)-\wp^{(1/3)}(u-2v)]
   -x(3v)\,y^{(1/3)}_t(u-2v)\nonumber\\
   && + y^{(1/3)}_t(2u-v)\,x(-3u)
   -2x_d(3u)\,y^{(1/3)}_t(-u-v)+2y^{(1/3)}_t(u+v)\,x_d(-3v)
   \nonumber\\
   && - 3x^{(1/3)}_t(2u-v)\,y(-3u)+3y(3v)\,x^{(1/3)}_t(u-2v)
   -3x^{(1/3)}_t(u+v)\,y_d(-3v)\nonumber\\
   &&\hspace*{4.5cm}+\  3y_d(3u)\,x^{(1/3)}_t(-u-v)=0,
   \label{twtripiden}
\end{eqnarray}
\begin{eqnarray}
   &&x^{(1/3)}(u-v)[\wp(2u-v)-\wp(u-2v)]
   -x^{(1/3)}(v)\,y(u-2v)\nonumber\\
   && + y(2u-v)\,x^{(1/3)}(-u)
   -2x^{(1/3)}_d(u)\,y(-u-v)+2y(u+v)\,x^{(1/3)}_d(-v)
   \nonumber\\
   && - x(2u-v)\,y^{(1/3)}(-u)+y^{(1/3)}(v)\,x(u-2v)
   -x(u+v)\,y^{(1/3)}_d(-v)\nonumber\\
   &&\hspace*{4.5cm}+\  y^{(1/3)}_d(u)\,x(-u-v)=0,
   \label{twtripiden2}
\end{eqnarray}
in which (\ref{twtripiden}) is obtained from the untwisted
third sum rule (\ref{tripiden}) by changing the triple
root functions \(x_t,y_t,z_t\) into
\(x_t^{(1/3)},y_t^{(1/3)},z_t^{(1/3)}\). Obviously this is
related to the two dimensional root system \(G_2\)
and its long root -long root pattern (\ref{g2longlong}).
This is responsible for the consistency of twisted \(G_2\) Lax pair
based on long roots discussed in \ref{g2long}.
The next twisted third sum rule (\ref{twtripiden2}) plays a role
in twisted \(G_2\) Lax pair
based on short roots discussed in \ref{g2short}.

\subsection{Twisted functions}
Here we give a simple set of functions with spectral
parameter satisfying all of the sum rules given in section 3,
(\ref{eq:ident1}), (\ref{eq:xydiden}), (\ref{tripiden}) and
section 5, (\ref{eq:ident1n})--(\ref{eq:twsecsum3}),
(\ref{eq:xydiden-n}), (\ref{twtripiden}) and (\ref{twtripiden2}):
\begin{eqnarray}
   x(u,\xi) &=& {\sigma(\xi-u)\over {\sigma(\xi)\sigma(u)}},
   \quad
   x_d(u,\xi) = x(u,2\xi), \quad
   x_t(u,\xi) = x(u,3\xi),
   \label{sol-1}\\
   x^{(1/2)}(u,\xi) &=& {\sigma^{(1/2)}(\xi/2-u)\over
   {\sigma^{(1/2)}(\xi/2)\sigma^{(1/2)}(u)}}
   \exp\left[e_{1}\xi\,u/2\right]\quad
   x_d^{(1/2)}(u,\xi) = x^{(1/2)}(u,2\xi),
   \label{sol-2}\\
   x^{(1/3)}(u,\xi) &=& {\sigma^{(1/3)}(\xi/3-u)\over
   {\sigma^{(1/3)}(\xi/3)\sigma^{(1/3)}(u)}}
   \exp\left[(2/3)\wp(2\omega_{1}/3)\,\xi\,u\right], \nonumber \\
   x_d^{(1/3)}(u,\xi) &=& x^{(1/3)}(u,2\xi), \qquad \quad
   x_t^{(1/3)}(u,\xi) = x^{(1/3)}(u,3\xi),
   \label{sol-3}\\
   x^{(1/4)}(u,\xi) &=& {\sigma^{(1/4)}(\xi/4-u)\over
   {\sigma^{(1/4)}(\xi/4)\sigma^{(1/4)}(u)}}
   \exp\left[(e_{1}/4+\wp(\omega_{1}/2)/2)\,\xi\,u\right],
   \nonumber\\
   x_d^{(1/4)}(u,\xi) &=& x^{(1/4)}(u,2\xi),
   \label{sol-4}
\end{eqnarray}
in which the superscript $(1/n)$ on the Weierstrass sigma
functions denotes that their primitive periods are
$\{2\omega_{1}/n,2\omega_{3}\}$
\[
   \sigma^{(1/n)}(u)=\sigma(u|2\omega_{1}/n,2\omega_{3})
\]
and those without them have the
usual periods
$\{2\omega_{1},2\omega_{3}\}$, {\it i.e.}, have $n=1$.
In order to show that they actually satisfy the sum rules,
the monodromy property of these functions is important.
The minimum intervals and corresponding monodromies
in $u$ and $\xi$
for these functions are given in the following table:
\begin{center}
 \begin{tabular}{|c|c|c|c|c|}
  \hline
  &\multicolumn{2}{|c|}{$u$}&\multicolumn{2}{|c|}
  {$\xi$}\\[4pt]\cline{2-5}
  {function} & interval & monodromy & interval & monodromy \\[4pt]
  \hline
  \raisebox{-2.0ex}[0cm][0cm]{$x^{(1/n)}(u,\xi)$} & $2\omega_{1}/n$ &
  $\exp[-(2/n)\eta_{1}\xi]$ & $2\omega_{1}$ &
  $\exp[-2\eta_{1}u]$ \\[4pt]
  \cline{2-5} & $2\omega_{j}$ & $\exp[-2\eta_{j}\xi]$
  & $2n\omega_{j}$ &
  $\exp[-2n\eta_{j}u]$ \\[4pt] \hline
  \raisebox{-2.0ex}[0cm][0cm]{$x_d^{(1/n)}(u,\xi)$} & $2\omega_{1}/n$ &
  $\exp[-(4/n)\eta_{1}\xi]$&
  $\omega_{1}$ & $\exp[-2\eta_{1}u]$ \\[4pt] \cline{2-5}
  & $2\omega_{j}$ & $\exp[-4\eta_{j}\xi]$ & $n\omega_{j}$ &
  $\exp[-2n\eta_{j}u]$ \\[4pt] \hline
  \raisebox{-2.0ex}[0cm][0cm]{$x_t^{(1/n)}(u,\xi)$} & $2\omega_{1}/n$ &
  $\exp[-(6/n)\eta_{1}\xi]$ &
  $2\omega_{1}/3$ & $\exp[-2\eta_{1}u]$ \\[4pt] \cline{2-5}
  & $2\omega_{j}$ & $\exp[-6\eta_{j}\xi]$ & $2n\omega_{j}/3$
  &
  $\exp[-2n\eta_{j}u]$ \\[4pt] \hline
 \end{tabular}
\end{center}
\begin{center}
Table 1: Monodromy Properties
\end{center}

This means that, for example,
\begin{eqnarray*}
 x^{(1/n)}(u+2\omega_{1}/n,\xi)&=&\exp[-(2/n)\eta_{1}\xi]\,
 x^{(1/n)}(u,\xi),\\
 x^{(1/n)}(u,\xi+2\omega_{1})&=&\exp[-2\eta_{1}u]\,
 x^{(1/n)}(u,\xi),
 \quad
 \eta_j=\zeta(\omega_j),\quad j=1,2,3.
\end{eqnarray*}

The monodromies of the functions $x^{(1/n)}$ for $n=2,3,4$ may be
derived  from the monodromy
of the function $x$ by using the following relations
\begin{eqnarray}
 x^{(1/2)}(u,\xi) &=& {x(u,\xi/2)\,x(u+\omega_{1},\xi/2)\over
 x(\omega_{1},\xi/2)}, \label{half-x}\\
 x^{(1/3)}(u,\xi) &=&
 {x(u,\xi/3)\,x(u+2\omega_{1}/3,\xi/3)\,x(u+4\omega_{1}/3,\xi/3)
 \over{x(2\omega_{1}/3,\xi/3)\,x(4\omega_{1}/3,\xi/3)}},
 \label{third-x}\\
 x^{(1/4)}(u,\xi) &=&
 {x(u,\xi/4)\,x(u+\omega_{1}/2,\xi/4)\,x(u+\omega_{1},\xi/4)
 \,x(u+3\omega_{1}/2,\xi/4)\over{x(\omega_{1}/2,\xi/4)
 \,x(\omega_{1},\xi/4)\,x(3\omega_{1}/2,\xi/4)}}.
 \label{fourth-x}
\end{eqnarray}

The following factorisation identities of these solutions
are useful (the dependence on \(\xi\) on the l.h.s. is
suppressed for brevity):
\begin{eqnarray}
 x(u)\,x(-u) &=& -\wp(u) + \wp(\xi), \label{FactorRules}\\
 \nonumber x^{(1/n)}(u)\,x^{(1/n)}(-u) &=& -\wp^{(1/n)}(u) +
 \wp^{(1/n)}(\xi/n),\quad n=2,3,4, \\
 \nonumber
 x_d^{(1/n)}(u)\,x_d^{(1/n)}(-u) &=& -\wp^{(1/n)}(u) +
 \wp^{(1/n)}(2\xi/n), \quad n=2,3,4,\\
 \nonumber
 x_t^{(1/3)}(u)\,x_t^{(1/3)}(-u) &=& -\wp^{(1/3)}(u) +
 \wp^{(1/3)}(\xi),
\end{eqnarray}
which give the potentials at various periods and are direct
consequences of the first sum rule, as explained in
(\ref{eq:facform}). The second group is responsible for the
renormalised constants in the extended twisted models
(\(n=1,2\)):
\begin{eqnarray}
 x^{(1/n)}(2u)x_d^{(1/2n)}(-u)+x^{(1/n)}(-2u)x_d^{(1/2n)}(u)
 &=&-\wp^{(1/2n)}(u)+
 \wp^{(1/n)}(\xi/n)\nonumber\\
 &&\hspace*{0.5cm}\ \ -\wp^{(1/n)}(\omega_1/n),
 \label{fac1d1}\\
 x^{(1/n)}(u)x^{(1/2n)}(-u)+x^{(1/n)}(-u)x^{(1/2n)}(u)
 &=&-2\wp^{(1/n)}(u)+
 \wp^{(1/2n)}(\xi/2n)\nonumber\\
 &&\hspace*{0.5cm}\ \ +\wp^{(1/n)}(\omega_1/n),
 \label{fac11n}\\
 x_d^{(1/n)}(u)x_d^{(1/2n)}(-u)+x_d^{(1/n)}(-u)x_d^{(1/2n)}(u)
 &=&-2\wp^{(1/n)}(u)+
 \wp^{(1/2n)}(\xi/n)\nonumber\\
 &&\hspace*{0.5cm}\ \ +\wp^{(1/n)}(\omega_1/n),
 \label{fac1d1dn}\\
 x(2u)\,x_d^{(1/4)}(-u)+x(-2u)\,x_d^{(1/4)}(u)&=&
 -\wp^{(1/2)}(u) +
 \wp^{(1/2)}(\xi/2). \label{fac-14d}
\end{eqnarray}
They are obtained from the twisted second sum rules,
(\ref{fac1d1}) from
(\ref{eq:twsecsum1}), (\ref{fac11n}) from
(\ref{eq:twsecsum2}) and
(\ref{fac1d1dn}) from (\ref{eq:twsecsum3}).
Likewise we obtain the following relations from the
twisted third sum rules, (\ref{twtripiden}) and
(\ref{twtripiden2}):
\begin{eqnarray}
 x(3u)x_t^{(1/3)}(-u)&+&x(-3u)x_t^{(1/3)}(u)
 + x_d(3u)x_t^{(1/3)}(-2u)+x_d(-3u)x_t^{(1/3)}(2u)\nonumber\\
 &=&-\wp^{(1/3)}(u)+3\wp(\xi) -2\wp(2\omega_{1}/3),
 \label{fac1t3}\\
 x(u)x^{(1/3)}(-u)&+&x(-u)x^{(1/3)}(u)
 + x(2u)x_d^{(1/3)}(-u)+ x(-2u)x_d^{(1/3)}(u)\nonumber\\
 &=&-3\wp(u)+\wp^{(1/3)}(\xi/3) + 2\wp(2\omega_{1}/3).
 \label{fac113}
\end{eqnarray}

\subsection{Degenerate cases}
Twisted Calogero-Moser models  can also be defined for
the trigonometric and hyperbolic potentials. Since these
potential functions have only one period, the half or
third-period potentials are simply obtained as:
\[
    \cot^2u,\quad \coth^2 u \Rightarrow
    \left\{
     \begin{array}{ccc}
   4\cot^2 2u,& 4\coth^2 2u&\\
   9\cot^2 3u,& 9\coth^2 3u& \mbox{etc.}
    \end{array}
  \right.
\]
The Lax pairs with spectral parameter for these potentials
can be obtained from those Lax pairs given in section 4
by taking the degenerate limits of the functions
\(\{x,x^{(1/2)},x_d,x_d^{(1/2)},...\}\).
With the help of the formulas (\ref{half-x})--(\ref{fourth-x}) the
degenerate limits of the twisted functions
\(\{x^{(1/2)},x^{(1/3)},x_d^{(1/2)},x_d^{(1/3)},...\}\)
can be obtained from those of the
untwisted functions
\(\{x,,x_d,x_t,...\}\)  given in section \ref{untwistdeg}.

\subsubsection{Imaginary period infinite: Trigonometric potential}
In this case, as we have seen, the untwisted functions
(\ref{untwiste-sols})
\[
 x(u,\xi)={\sigma(\xi-u)\over{\sigma(\xi)\sigma(u)}},
    \quad
   x_d(u,\xi)={\sigma(2\xi-u)\over{\sigma(2\xi)\sigma(u)}}
   \quad
   x_t(u,\xi)={\sigma(3\xi-u)\over{\sigma(3\xi)\sigma(u)}},
\]
reduce to (\ref{imainffunx})
\[
   x(u,\xi)=\kappa\left(\cot \kappa u-\cot \kappa
   \xi\right)\,e^{-a\xi u},
   \quad
   x_d(u,\xi)=x(u,2\xi),
   \quad
   x_t(u,\xi)=x(u,3\xi) \quad \kappa=\sqrt{12a}.
\]
\noindent From these we obtain, as expected,
\begin{eqnarray}
   x^{(1/n)}(u,\xi)&=&n\kappa\left(\cot n\kappa u-\cot \kappa
   \xi\right)\,e^{-a\xi u},
   \quad  \quad
n=2,3,4,  \nonumber\\
x_d^{(1/n)}(u,\xi)&=&x^{(1/n)}(u,2\xi),
\qquad \qquad x_t^{(1/3)}(u,\xi)=x^{(1/3)}(u,3\xi).
\label{twtrig}
\end{eqnarray}
In deriving these formulas from (\ref{half-x})--(\ref{fourth-x})
use is made of an interesting trigonometric identity:
\begin{equation}
 \left(\cot u-\cot v\right)\prod_{j=1}^{n-1}
 {\left(\cot(u+{j\pi\over
 n})-\cot v\right)\over{\left(\cot({j\pi\over
 n})-\cot v\right)}}=n\left(\cot nu-\cot nv\right),
 \quad n=1,2,\ldots,.
\end{equation}
The spectral parameter independent functions are
\begin{eqnarray}
   x(u)&=&x_d(u)=x_t(u)=a\,(\cot a u\mp1),\quad
   x^{(1/2)}(u)=x_d^{(1/2)}(u)=2a\,(\cot 2a u\mp1),
   \nonumber \\
   x^{(1/3)}(u)&=&x_d^{(1/3)}(u)=x_t^{(1/3)}(u)=3a\,(\cot 3a u\mp1),
   \nonumber \\
   x^{(1/4)}(u)&=&x_d^{(1/4)}(u)=4a\,(\cot 4a u\mp1),
   \quad a: const.
   \label{degcot}
\end{eqnarray}

\subsubsection{Real period infinite: Hyperbolic potential}
In this degenerate limit, twisting along the direction
\(\{2\omega_1,2\omega_3\}\to \{2\omega_1/n,2\omega_3\}\)
gives trivial results:
\begin{equation}
 x(u)=x^{(1/2)}(u)=x^{(1/3)}(u),\quad
 x_d(u)=x_d^{(1/2)}(u), \ldots,
\end{equation}
since \(\omega_1=\infty\) (\ref{realinf}).
That is, the twisted models are identical with the
untwisted models.
 Twisting along other directions
give non-trivial results. For example,
\[
   \{2\omega_1,2\omega_3\}\to \{2\omega_1,2\omega_3/n\},
\]
the degenerate limit of the functions can be obtained in a
similar way as above:
\[
   x(u,\xi)=\kappa\left(\coth \kappa u-\coth \kappa
   \xi\right)\,e^{a\xi u},
   \quad
   x_d(u,\xi)=x(u,2\xi),
   \quad
   x_t(u,\xi)=x(u,3\xi) \quad \kappa=\sqrt{12a}.
\]
\noindent From these we obtain, as above,
\begin{eqnarray}
   x^{(1/n)}(u,\xi)&=&n\kappa\left(\coth n\kappa u-\coth \kappa
   \xi\right)\,e^{a\xi u},
   \quad  \quad
n=2,3,4,  \nonumber\\
x_d^{(1/n)}(u,\xi)&=&x^{(1/n)}(u,2\xi),
\qquad \qquad x_t^{(1/3)}(u,\xi)=x^{(1/3)}(u,3\xi).
   \label{twhyp}
\end{eqnarray}
The spectral parameter independent functions are
\begin{eqnarray}
   x(u)&=&x_d(u)=x_t(u)=a\,(\coth a u\pm1),\quad
   x^{(1/2)}(u)=x_d^{(1/2)}(u)=2a\,(\coth 2a u\pm1),
   \nonumber \\
    x^{(1/3)}(u)&=&x_d^{(1/3)}(u)=x_t^{(1/3)}(u)=3a\,(\coth 3a
   u\pm1),
   \nonumber \\
   x^{(1/4)}(u)&=&x_d^{(1/4)}(u)=4a\,(\coth 4a u\pm1),
   \quad a: const.
   \label{degcoth}
\end{eqnarray}
\section{Summary and comments}
The construction of universal Lax pairs for all of
the Calogero-Moser models based on root systems is completed.
This paper presents  universal Lax pairs for the twisted models
based on non-simply laced root systems, following those for the
models based on simply laced root systems given in the first paper
\cite{bcs} and those for the untwisted models
based on non-simply laced root systems constructed in the second
paper \cite{bst}.
All of the Lax pairs for the twisted models presented here are new,
except for the one for the \(F_{4}\) model based on the short roots.
These Lax pairs are for the elliptic potentials and  contain
a spectral parameter.
The explicit forms and the properties of the functions
appearing in the Lax pairs,
the untwisted and twisted functions,  are determined from
the functional equations (which we call sum rules)
necessary and sufficient for the
consistency of the Lax pairs.

The spectral parameter dependent Lax pairs for the hyperbolic,
trigonometric and rational potentials for the untwisted as well as
the twisted models are derived as the degenerate limits of the
elliptic potential case.
These, we believe, have  not been reported yet.
All of the Lax pairs discussed in this paper are of the root type
and contain at least as many independent coupling constants
as the number of independent Weyl orbits in the root systems. Our
main result is very simple. The Lax pairs for the twisted models
can be obtained  from those for the untwisted models (given in
paper II) by replacing the untwisted functions
\(\{x,y,z\), \(x_d,y_d,z_d\), \(x_t,y_t,z_t\}\) by the twisted
functions \(\{x^{(1/n)},y^{(1/n)},z^{(1/n)}\),
\(x^{(1/n)}_d,y^{(1/n)}_d,z^{(1/n)}_d\),
\(x^{(1/n)}_t,y^{(1/n)}_t,z^{(1/n)}_t\}\).
This also applies to the other type of universal Lax pairs.
The minimal type Lax pairs can be obtained simply from those for the
untwisted models by replacing the untwisted functions \(\{x,y,z\}\)
by the twisted functions \(\{x^{(1/2)},y^{(1/2)},z^{(1/2)}\}\)
(for the \(G_{2}\) model, \(\{x^{(1/3)},y^{(1/3)},z^{(1/3)}\}\))
for the short roots.

As for the twisted models based on \(B_{n}\), \(C_{n}\) and
\(BC_{n}\) root systems, a new type of potential terms with
independent coupling constants can be added without destroying
integrability.
They are called extended twisted models and  have three, three and
five independent coupling constants for \(B_{n}\), \(C_{n}\) and
\(BC_{n}\) models, respectively.
The Lax pairs for the twisted \(G_{2}\) model have new features.
This  model has an  elliptic potential function with a pair of
primitive periods
\(\{2\omega_{1},2\omega_{3}\}\) for the long roots and another
elliptic  potential function with third-period
\(\{2\omega_{1}/3,2\omega_{3}\}\)  for the short roots. New sum
rules appear in the proof of the consistency of the Lax pairs
for this model.

\begin{center}
{\bf ACKNOWLEDGMENTS}
\end{center}
We thank E. Corrigan, M. Olshanetsky and K. Takasaki for useful discussion
and comments.
This work is partially supported  by the Grant-in-aid from the
Ministry of Education, Science and Culture, priority area
(\#707)  ``Supersymmetry and unified theory of elementary
particles". A.\,J.\,B is supported by the Japan Society for the
Promotion of Science  and the National Science Foundation under
grant no. 9703595.


\end{document}